\documentclass[pra,twocolumn,showpacs,a4paper]{revtex4}
\usepackage{color}
\usepackage{epsfig}
\usepackage{amsmath}
\usepackage{graphicx}
\usepackage{bm}

\newcommand{\cav}{{\rm c}}
\newcommand{\done}{\Delta_1}
\newcommand{\aone}{a_1}
\newcommand{\aoned}{a_1^{\dagger}}
\newcommand{\dtwo}{\Delta_2}
\newcommand{\atwo}{a_2}
\newcommand{\atwod}{a_2^{\dagger}}

\newcommand{\di}{\Delta_i}
\newcommand{\ai}{a_i}
\newcommand{\aid}{a_i^{\dagger}}

\newcommand{\Fi}{F_i}
\newcommand{\Fid}{F_i^{\dag}}
\newcommand{\Fone}{F_1}
\newcommand{\Foned}{F_1^{\dag}}
\newcommand{\Ftwo}{F_2}
\newcommand{\Ftwod}{F_2^{\dag}}

\newcommand{\mec}{{\rm m}}
\newcommand{\omm}{\omega_\mec}
\newcommand{\am}{a_\mec}
\newcommand{\amd}{a_\mec^{\dagger}}
\newcommand{\Xm}{x_\mec}
\newcommand{\Xmbar}{\bar{x}_\mec}
\newcommand{\lm}{\ell_\mec}
\newcommand{\gm}{\gamma_\mec}
\newcommand{\nm}{\bar{n}_\mec}
\newcommand{\Gm}{g_\mec}

\newcommand{\at}{{\rm at}}
\newcommand{\oma}{\omega_\at}
\newcommand{\aat}{a_\at}
\newcommand{\aad}{a_\at^{\dagger}}
\newcommand{\Xa}{x_\at}
\newcommand{\Xabar}{\bar{x}_\at}
\newcommand{\la}{\ell_\at}
\newcommand{\ga}{\Gamma_\at}

\newcommand{\Ga}{g_\at}

\newcommand{\be}{\begin{equation}}
\newcommand{\ee}{\end{equation}}
\newcommand{\hc}{\mathrm{h.c.}}

\newcommand{\Geff}{G}
\newcommand{\U}{U_0}
\newcommand{\hpi}{h_{+,i}}
\newcommand{\hmi}{h_{-,i}}

\newcommand{\ts}{t_{\rm s}}

\begin{document}

\title{Single-Atom Cavity QED and Opto-Micromechanics}

\author{M. Wallquist, K. Hammerer, P. Zoller}
\address{Institute for Theoretical Physics, University of Innsbruck, and
Institute for Quantum Optics and Quantum Communication, Austrian Academy of Sciences,
Technikerstrasse 25, 6020 Innsbruck, Austria}
\address{Norman Bridge Laboratory of Physics 12-33, California Institute of Technology, Pasadena, CA 91125 USA}
\author{C. Genes}
\address{Institute for Theoretical Physics, University of Innsbruck, and
Institute for Quantum Optics and Quantum Communication, Austrian Academy of Sciences,
Technikerstrasse 25, 6020 Innsbruck, Austria}
\author{M. Ludwig, F. Marquardt}
\address{Department of Physics, Center for NanoScience, and Arnold Sommerfeld Center for Theoretical Physics, Ludwig-Maximilians-Universit\"at M\"unchen, Theresienstr. 37, D-80333 Munich, Germany }
\author{P. Treutlein}
\address{Max-Planck-Institute of Quantum Optics and Department of Physics, Ludwig-Maximilians-Universit\"at M\"unchen, Schellingstr. 4, D-80799 Munich, Germany }
\author{J. Ye}
\address{JILA, National Institute of Standards and Technology and University of Colorado, Boulder, CO 80309-0440 USA}
\address{Norman Bridge Laboratory of Physics 12-33, California Institute of Technology, Pasadena, CA 91125 USA}
\author{H. J. Kimble}
\address{Norman Bridge Laboratory of Physics 12-33, California Institute of Technology, Pasadena, CA 91125 USA}

\begin{abstract}
In a recent publication \cite{hammerer_strong_2009}
we have shown the possibility to achieve strong coupling of the quantized motion of a micron-sized mechanical system to the motion of a single trapped atom. In the proposed setup the coherent coupling between a SiN membrane and a single atom is mediated by the field of a high finesse cavity, and can be much larger than the relevant decoherence rates. This makes the well-developed tools of CQED (cavity quantum electrodynamics) with single atoms available in the realm of cavity optomechanics. In this paper we elaborate on this scheme and provide detailed derivations and technical comments. Moreover, we give numerical as well as analytical results for a number of possible applications for transfer of squeezed or Fock states from atom to membrane as well as entanglement generation, taking full account of dissipation. In the limit of strong-coupling the preparation and verification of non-classical states of a mesoscopic mechanical  system is within reach.
\end{abstract}

\maketitle
\section{Introduction}

The quantum regime of optomechanical systems
\cite{Kippenberg_review,Marquardt_review_2009} -- in particular micro or
nanomechanical oscillators coupled to the optical field in a cavity --
has recently received considerable attention, mainly owing to the
experimental progress in quantum ground state cooling
\cite{groblacher_demonstration_2009,schliesser_resolved_2009} and
strong coupling dynamics \cite{Marquardt_PRL2007,dobrindt_parametric_2008,wilson_rae_cavity_2008,groeblacher_observation_2009,huang_normal_2009}.
Combining opto-micromechanics with low-loss dielectric membranes \cite{ThompsonNature2008,wilson_cavity_2009} on the one hand, with cavity QED \cite{miller_trapped_2005} with single or many atoms on the other hand, a hybrid system
emerges that can be a testbed for experiments on coherent dynamics
between microscopic (single atom or ensemble of atoms)
and macroscopic (micro-mechanical oscillator) systems. Given the
already well-developed toolbox for the manipulation of atomic states such an interface
can be used for indirect preparation and manipulation of quantum
states of mesoscopic mechanical oscillators. Moreover, in view of applications such as quantum information processing, it seems timely to ask for quantum hybrid systems which combine the advantages of physically different systems, each with a unique set of properties and capabilities, in a compatible experimental setup. A hybrid atomic-mechanical system would be one such example.

A few recent theoretical proposals advance the possibility of
coupling ensembles of atoms to mechanical resonators. Most generally
the interaction is mediated by a light field that couples the
mechanical resonator via the radiation pressure effect to either
internal levels of the atoms \cite{genes_emergence_2008,ian_cavity_2008,hammerer_establishing_2009}, or to their motional degrees of freedom \cite{meiser_coupled_2006}, which can result e.g. in cooling of the mechanical resonator via a bath of atoms \cite{genes_micro_2009}.
Also a direct coupling has been proposed where a magnetic tip mounted on a cantilever provides a Zeeman coupling to the atomic spin of the Bose-condensed \cite{treutlein_bose_2007} or ultracold \cite{geraci_ultracold_2009} atoms. A number of proposals discuss the possibility to couple the motion of a microresonator to single two level systems, realized e.g. in a quantum dot \cite{WilsonRaePRL2004}, in a
nitrogen-vacancy impurity in diamond \cite{rabl_strong_2009}, or in superconducting circuits such as a Cooper-pair box \cite{armour_entanglement_2002,Jaehne_NJP2008}, a SQUID \cite{ZhouPRL2006,BuksPRB2006} or a flux qubit \cite{XueNJP2007}.

The direct coupling of the motion of a single microscopic body such as a single atom to a macroscopic mechanical oscillator is considerably more challenging. Typically, the interaction strength is governed by a small intrinsic parameter which scales as $\sqrt{m/M}\sim10^{-7}-10^{-4}$, where $m$ and $M$ are the masses of the atom and the mechanical oscillator. This is true e.g in \cite{tian_coupled_2004} where the motion of an ion in a trap is coupled to the vibrations of nano-electrodes providing the trap potential. An alternative route is however possible, where an indirect cavity-mediated coupling circumvents the limitations
imposed by the small mass ratio, as presented in our recent proposal \cite{hammerer_strong_2009}. Thereby
a strong coupling is achievable between a single trapped atom and the
motion of a membrane, where the coupling strength can exceed the dissipative rates by a
factor of ten for present or near future experimental parameters.

In this article we elaborate on the mechanism described in our previous letter
\cite{hammerer_strong_2009}, and provide more details and applications of the scheme. The paper is structured as follows. Section~\ref{Sec:Overview} presents an overview and qualitative picture of our results. In Sec.~\ref{Sec:effMA} the reduced master equation describing the cavity-mediated membrane-atom interaction is derived in detail, and results are presented in particular for the dispersive regime. Sec.~\ref{sec:cohevol} specializes on the regime of strong membrane-atom coupling, and examples of state transfer are presented. In addition, we describe how to produce entanglement between atom and membrane by modulating the input laser intensity in time, leading to a two-mode squeezing Hamiltonian. Sec.~\ref{Sec:Techdet} discusses technical details regarding the specific setup that we have in mind, and finally we discuss the result and conclude in Sec.~\ref{Sec:Disc}. Mathematical details of the derivation are presented in Appendices.

\section{Overview}
\label{Sec:Overview}

\begin{figure}[t]
\begin{centering}
\includegraphics[width=\columnwidth]{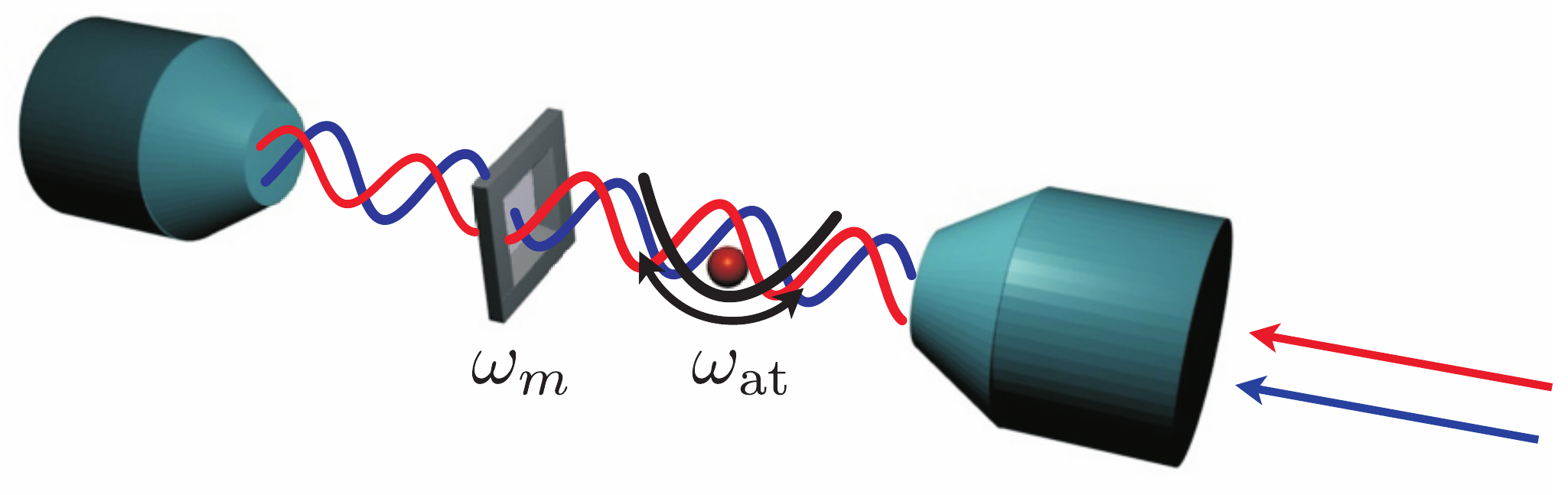}
\caption{Dynamic intracavity field provides strong interface between the motion of a single trapped atom and the vibrations of a micron-sized membrane.}
\label{figSetup}
\end{centering}
\end{figure}

In the setup proposed in \cite{hammerer_strong_2009}, the recent development within micromechanics with membranes in optical cavities \cite{ThompsonNature2008} is combined with single trapped atom cavity QED \cite{miller_trapped_2005}. As shown in Fig.~\ref{figSetup}, we consider an optomechanical system where a micron-sized dielectric membrane is placed in a laser driven high-finesse cavity and coupled through radiation pressure to the cavity field quadratures, with the coupling strength controlled by the laser power through the intracavity amplitude. In this setup, the membrane vibration manifests itself as a dynamic detuning of the driven cavity modes. For a cavity mode driven by a laser detuned from the cavity resonance this dynamic detuning translates into a dynamic intracavity field intensity. If now a single atom is trapped in the optical dipole potential provided by the cavity field, the membrane vibration couples via the dynamics of the optical trap to the motion of the atom, and vice versa. This coupling is strongly enhanced by a large steady-state field amplitude and the cavity finesse, which is a key ingredient in achieving the strong coupling regime.

\subsection{Effective Master Equation and Strong Coupling}

The focus of our analysis is a configuration where the cavity field serves merely as a quantum bus and can be effectively eliminated from the dynamics, giving rise to a coupled oscillator dynamics for the reduced system comprising the membrane and the atom [$\hbar=1$],
\be
H = \omm \amd\am + \oma \aad\aat -  G (\am + \amd)(\aat + \aad).
\label{Eq:Hatmec}
\ee
In this Hamiltonian the first and second terms describe the bare micromechanical oscillator and harmonic motion of the trapped atom, respectively, with $\am$ ($\aat$) being the mechanical (atomic motion) annihilation operator. $\omm$ and $\oma$ are the respective oscillation frequencies. The linear form of this interaction would provide a quantum interface for coherent transfer of quantum states between the mechanical oscillator and the atom, opening the door to coherent manipulation, preparation and measurement of micromechanical objects via well-developed tools of atomic physics, as will be detailed in Sec.~\ref{sec:cohevol}.

However, the cavity mediated, coherent dynamics will compete with a number of dissipative processes, such that the full dynamics will be described by a master equation,
\be
\dot{\rho} = -i [H,\rho] + L_\mec (\rho) + L_\at (\rho)+ L_\cav (\rho).
\label{Eq:MEeff}
\ee
The three Liouvillian terms describe the respective sources of dissipation, with $L_\mec$ including the thermal heating of the membrane vibration and $L_\at$ including the atomic momentum diffusion due to spontaneous emission. Furthermore, a cavity-mediated coupling comes naturally at the price of cavity-induced decoherence via photon leakage, $L_\cav$. Our goal here is to construct a setup obeying the master equation (\ref{Eq:MEeff}) with a Hamiltonian term \eqref{Eq:Hatmec} where the interaction between the atom and the membrane is \textit{resonant}, i.e. $\omm\simeq\oma$, and \textit{strong}, i.e. the coupling constant $\Geff$ is larger than the relevant decoherence rates $\Gamma_\cav,\Gamma_\mec,\Gamma_\at$ corresponding to the dissipative processes described by $L_\cav,L_\mec,L_\at$, respectively. In fact, we will show that for state of the art experimental parameters small ratios $(\Gamma_\cav,\Gamma_\mec,\Gamma_\at)/\Geff \simeq 0.1$ are within reach.

\subsection{Qualitative Picture of Linear Coupling}

\begin{figure}[t]
\begin{centering}
\includegraphics[width=\columnwidth]{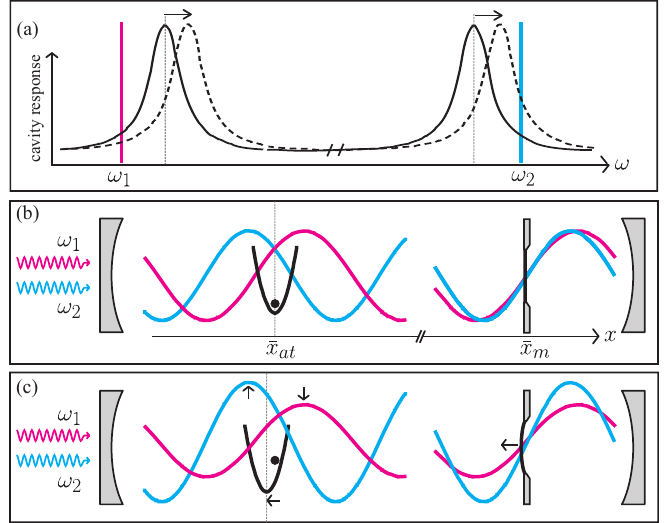}
\caption{Linear atom-membrane coupling mediated by two driven cavity modes.
(a) One mode is driven on the red side, the other on the blue side. When the mode frequencies shift due to the membrane vibration (dashed line), the cavity response is reduced for one mode and enhanced for the other.
(b) Atom and membrane in equilibrium inside the driven cavity.
(c) When the membrane vibrates around its equilibrium, the oppositely changing cavity response for the respective modes shifts the equilibrium of the combined atom potential. }
\label{figMechanism}
\end{centering}
\end{figure}

A strong linear coupling as described in Eq. (\ref{Eq:Hatmec}) is obtained in a configuration involving {\it two} driven cavity modes of frequencies $\omega_{\cav,1}$ and $\omega_{\cav,2}$, as shown in Fig.~\ref{figMechanism}(a). The two modes are driven by lasers of frequencies $\omega_1$ and $\omega_2$, respectively, where the first laser is tuned to the red side of its cavity resonance, $\omega_1 - \omega_{\cav,1} < 0$, and the second laser is tuned to the blue side, $\omega_2 - \omega_{\cav,2} > 0$. By a proper choice of cavity modes, and with an internal structure of the specific atom as shown in Fig.~\ref{figAtomInt}, both lasers separately provide red-detuned optical lattices, which combine into a potential where we trap a single atom in one of the wells, see Fig.~\ref{figMechanism}(b). With wave vectors $k_1 \neq k_2 $ and assuming equally large intracavity amplitudes, the two lattices have opposite slopes at the equilibrium position $\Xabar$ of the atom. Moreover, the particular well is chosen such that the slopes are close to maximal, such that the response of the cavity field amplitude to the atomic motion is close to maximal. Similarly, the membrane is positioned at $\Xmbar$ half-way between a field node and anti-node, where the linear opto-mechanical coupling is maximal \cite{ThompsonNature2008}, see Fig.~\ref{figMechanism}(b). The position $\Xmbar$ is chosen such that both fields have similar slope (with the same sign), thus react equally to the membrane vibration. The displacement of the membrane thus shifts the cavity resonances, as shown by the dashed lines in Fig.~\ref{figMechanism}(a). With the two lasers being tuned to different sides of their respective resonances, during the membrane displacement one driving laser will come closer to resonance, with resulting enhanced intracavity field, and the other one farther off resonance with resulting reduced intracavity field. Consequently we will find that one of the atomic lattice potentials is getting deeper, the other one getting more shallow, as seen in Fig.~\ref{figMechanism}(c), thus shifting the atomic trapping potential. Due to this spatial shift being proportional to $\Xm$, the result is an overall $\sim \Xa\Xm$ coupling as in Eq. (\ref{Eq:Hatmec}).

With this construction, the cavity field can provide the leverage to couple two objects with mass ratio on the order of $10^{-13}$. Imagine for illustration that we were to achieve a similar coupling with a mechanical device like a seesaw: To balance the torques would require a lever ratio of the same order of magnitude; 15 mm on one side and the earth-sun distance on the other side.

\section{Model for Cavity Mediated Membrane--Atom Coupling}
\label{Sec:effMA}

After the qualitative description in the last section, we move on to a detailed presentation of the system consisting of a moving atom and a vibrating membrane coupled to driven cavity modes. Further, we will show how to obtain the reduced atom-membrane dynamics described by Eq. (\ref{Eq:MEeff}) by eliminating the cavity degrees of freedom, and finally we will identify the regime of strong coupling.

\subsection{Detailed Derivation of Effective Master Equation}
\label{Ssec:acmcint}

\subsubsection{Full master equation}

Our starting point is the complete master equation for the density operator $W$ describing the dynamics of cavity modes, atom and membrane motion,
\be
\dot W = -i  [H_{\rm sys} , W]
+ \left[ L_\mec + L_\at + \sum_i L_{i} \right] (W) ,
\label{Eq:MEW}
\ee
with the coherent dynamics contained in the system Hamiltonian $H_{\rm sys}$,
\[
H_{\rm sys} = H_{\rm motion} + H_\cav .
\]
Here $H_{\rm motion}$ takes into account the free harmonic motion of the membrane, modeled as a single-mode oscillator, and the kinetic energy of the atom with momentum $P_\at$,
\be
H_{\rm motion} = \omm\amd\am + P_\at^2 / 2m .
\label{Eq:Hmotionprime}
\ee
Further, $H_\cav$ contains the free cavity Hamiltonian, as well as the effect of the atomic motion and the membrane vibration on the cavity field. We will postpone its discussion to the next section, where we give the concrete form of $H_\cav$ for various setups, and here first address the remaining terms in the master equation (\ref{Eq:MEW}).

The Lindblad terms in the master equation (\ref{Eq:MEW}) describe dissipation of the membrane ($L_\mec$), the atom ($L_\at$) and the cavity modes ($L_i$) respectively. Here $i$ labels the cavity modes with photon annihilation operator $A_i$ obeying $[A_i , A_j^\dag]=\delta_{ij}$. Cavity decay at an amplitude decay rate $\kappa_i$ is described by,
\[
L_{i}(W) = \kappa_i {\cal D}[A_i](W),
\]
where we use the shorthand notation
\[
{\cal D}[a](W) = 2aW a^\dag - \{ a^\dag a , W \}_+
\]
for a Lindblad term with jump operator $a$.

\begin{figure}[t]
\begin{centering}
\includegraphics[width=0.3\columnwidth]{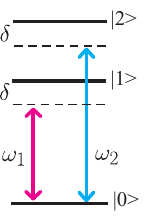}
\caption{Two lasers with frequencies $\omega_1$ and $\omega_2$ respectively drive two different internal atomic transitions with detunings $\delta$.}
\label{figAtomInt}
\end{centering}
\end{figure}
$L_\at$ describes the noise processes acting purely on the atoms, and could also include controlled dissipation such as Raman cooling. Here we assume each driven mode with wave number $k_i$ to couple to a different atomic transition $|0\rangle \leftrightarrow |i\rangle$, as sketched in Fig.~\ref{figAtomInt}, and focus on the photon recoil during spontaneous emission from the excited state $|i\rangle$ with rate $\gamma_i$ to the common ground-state $|0\rangle$. The effect of the photon recoil on the atomic motion is described by the following Lindblad term,
\begin{align}
& L_\at (W) = {1 \over 2} \sum_i \gamma_i s_{i} \nonumber \\
& \bigg(
2\int_{-1}^1 d\vartheta S_i(\vartheta) e^{i\vartheta k_i \Xa} \sqrt{u_i (\Xa)} W e^{- i\vartheta k_i \Xa}\sqrt{u_i (\Xa)}
\nonumber \\
& - \left\{ u_i (\Xa),W \right\}_+ \bigg)
\label{Eq:LatFull}
\end{align}
where $\gamma_i$ is the spontaneous emission rate and $s_{i}$ is the saturation parameter for transition $i$ \cite{CiracPRA1992}. $S_i (\vartheta)$ are (even) geometric functions whose exact expressions depend on the chosen transitions, and $u_i (\Xa)$ describe the spatial intensity profiles of the cavity modes with $\Xa$ the atomic position in the cavity. In the next section we will discuss this term in more detail in the Lamb-Dicke regime, where it simplifies considerably.

Finally, $L_\mec$ describes the membrane thermal contact via the finite temperature suspension, modeled as interaction with a thermal bath,
\be
L_\mec = {\gm \over 2} (\nm +1) {\cal D}[\am] + {\gm \over 2} \nm {\cal D}[\amd] ,
\label{Eq:Lm}
\ee
with $\gm$ the natural linewidth of the mechanical resonance, and $\nm$ its mean occupation in thermal equilibrium. The heating rate $\Gamma_\mec = \gm\nm \simeq k_B T/(\hbar Q_\mec)$ is related to the temperature $T$ of the contact and the mechanical quality factor $Q_\mec$. In addition to the thermal contact, we include membrane heating due to absorption of laser power. In fact, a fairly cautious estimate detailed in section \ref{Sec:Techdet} shows that with standard cryogenic precooling the natural lower limit for the temperature $T$ is set by light absorption within the membrane.

The model presented so far makes the following assumptions:
(1) atomic motion is accurately described by a 1D model, with the transverse confinement provided by the Gaussian intensity profile of the cavity fields, (2) internal atomic dynamics can be eliminated, assuming the laser drive to be sufficiently detuned from the atomic resonances (cf. Fig.~\ref{figAtomInt}), and (3) negligible internal coupling of the chosen membrane mode to vibrations of higher energy, allowing a single mode approximation.

\subsubsection{Linearization Around Equilibrium}

We will now proceed to discuss the cavity and interaction Hamiltonian $H_\cav$. Note first that despite the very different physical nature of the atom and the membrane, their effect on the cavity field can be collected in a unified description, where the cavity modes $A_i$ with frequencies $\omega_{\cav,i}$ see an index of refraction which depends on the respective positions $\Xa$ and $\Xm$ of the atom and the membrane along the cavity axis. This description assumes a Born-Oppenheimer type approximation, where the slow atom/membrane motion compared to the optical cavity frequencies allows a separation of timescales.

\paragraph*{Single Mode Setup:} We now first want to consider  a setup with only a single driven cavity mode, $i=1$, in order to illustrate a number of conceptual points. The generalization to the two-mode setup is then immediate. For a single mode the cavity Hamiltonian $H_\cav$ in the master equation \eqref{Eq:MEW}, taken in a frame rotating with the laser frequency $\omega$ (dropping the index $i$), is
\be
H_\cav = \left[ \omega_{\cav} \left( \Xa,\Xm \right) - \omega \right] A^\dag A
+ E \left( e^{i\phi} A^\dag  + {\rm h.c.} \right) .\label{Eq:Hcav1}
\ee
The first term is the cavity free energy in the rotating frame, which depends parametrically on the atom (membrane) position. The second term describes the laser drive of power $P$, such that $E = \sqrt{2P\kappa/\hbar\omega_{\cav}}$.

The strong drive field creates a steady-state intracavity field with amplitude $\alpha \gg 1$, which in turn provides a trap potential for the atom at a certain equilibrium point $\Xabar$ and mean force on the membrane, displacing it to a slightly shifted position $\Xmbar$. We are interested in the dynamics of the fluctuations of cavity amplitude and atom/membrane position around these equilibrium values. It is therefore convenient to move to a displaced frame where the dynamics is described by the fluctuations $a$ around the steady-state field,
\be
A = \alpha + a,\label{Eq:Exp0}
\ee
and the fluctuations $\delta\Xa$ and $\delta\Xm$ around the equilibrium atom and membrane positions,
\be
\Xa = \Xabar + \delta\Xa ,\qquad \Xm = \Xmbar + \delta\Xm .\label{Eq:Exp1}
\ee
Along this line, we expand the cavity mode frequency $\omega_{\cav} \left( \Xa,\Xm \right)$ around steady-state,
\begin{align}
\omega_{\cav}& \left( \Xa,\Xm \right)
 \simeq \omega_{\cav}^0
 + \left[\partial_\at
\omega_{\cav}\right] \delta\Xa + \left[\partial_\mec \omega_{\cav}\right]\delta\Xm
\label{Eq:cavdisp}\\
&+ {1 \over 2}
\left[\partial^2_\at \omega_{\cav}\right]\delta\Xa^2 +
{1 \over 2} \left[\partial^2_\mec \omega_{\cav} \right] \delta\Xm^2
\nonumber
+ \left[\partial^2_{\at,\mec} \omega_{\cav}\right]\delta\Xa\delta\Xm
\end{align}
with $\partial_\at$ and $\partial_\mec$ short for the partial derivative with respect to $\Xa$ and $\Xm$, evaluated at the atom and membrane equilibrium points. When the expansions \eqref{Eq:Exp0},\eqref{Eq:Exp1} and \eqref{Eq:cavdisp} are used in the full master equation \eqref{Eq:MEW} with Hamiltonian \eqref{Eq:Hcav1}, the steady state amplitude $\alpha$ and equilibrium positions of atom and membrane can be determined self-consistently by demanding that all terms vanish which are \textit{linear} in fluctuation operators $\delta\Xa,\,\delta\Xm$ and $a$. In particular one finds for the intra-cavity amplitude (see \cite{wilson_rae} for details),
\be
\alpha \simeq { E e^{i\phi} \over \left( \omega - \omega_\cav^0 \right) + i \kappa} .
\label{Eq:alpha}
\ee
The laser phase $\phi$ can be chosen for convenience such as to make $\alpha$ real.

In the resulting Hamiltonian all linear terms are thus systematically removed and the dynamics is governed by an effective Hamiltonian
\begin{align}
H_\cav\simeq&\, (\omega_{\cav}^0 -\omega)a^\dag a
+\frac{\alpha^2}{2}
\left[\partial^2_\at \omega_{\cav}\right]\delta\Xa^2
+\frac{\alpha^2}{2}\left[\partial^2_\mec \omega_{\cav} \right] \delta\Xm^2\nonumber\\
& + \alpha^2 \left[\partial^2_{\at,\mec}\omega_{\cav}\right]\delta\Xa\delta\Xm\nonumber\\
&+\alpha\left[\partial_\mec \omega_{\cav}\right]\delta\Xm(a+a^\dag)
+\frac{\alpha}{2}\left[\partial^2_\mec \omega_{\cav} \right] \delta\Xm^2(a+a^\dag)\nonumber\\
&+\frac{\alpha}{2}\left[\partial^2_\at \omega_{\cav} \right] \delta\Xa^2(a+a^\dag)
\label{Eq:LinHam},
\end{align}
where terms of fourth order in fluctuations (zeroth order in cavity amplitude) have been neglected.

In the first line we find the free energy of the cavity and the optical potential for the atom, providing a harmonic trap with a frequency determined by
$ m \oma^2 = \alpha^2 \left[\partial^2_\at \omega_{\cav}\right]$.
The corresponding term for the membrane provides a small correction to its mechanical frequency and can be neglected.

The term in the second line of Eq.~\eqref{Eq:LinHam} describes a \textit{direct} linear atom-membrane coupling of the form $\sim g_{\rm direct}(\aat + \aad)(\am + \amd)$,
where $\delta\Xa=\la (\aat + \aad)$ and $\delta\Xm=\lm (\am + \amd)$. The zero point fluctuations are given by,
\[
\la = \left({\hbar \over 2 m \oma}\right)^{1/2}, \qquad
\lm = \left({\hbar \over 2 M \omm}\right)^{1/2} ,
\]
and $\am$ now refers to the shifted frame for the membrane. Assuming that the cavity field provides the atomic trap as discussed above, with a trap frequency close to that of the membrane vibration, $\oma \sim \omm$, it can be checked easily that the direct coupling will be hampered by the small mass ratio,
\[
g_{\rm direct}\sim (\lm/\la)\oma \sim \sqrt{m / M}\oma ,
\]
thus would not reach the strong coupling regime for a single atom. We will see that the cavity-mediated, \textit{indirect} coupling can be many orders of magnitude larger, such that the direct coupling can be safely neglected in the following.

The third line in Eq.~\eqref{Eq:LinHam} describes a membrane-cavity interaction. As was discussed in detail in \cite{Jayich2008} a proper choice of the membrane position along the cavity axis can make either the first or the second term dominant. In the later case the cavity field couples to $\delta\Xm^2\sim(\am+\amd)^2$, which has interesting applications for measuring occupation numbers of the membrane. However this term is typically rather small as it scales like $\partial^2_\mec \omega_{\cav} \lm^2\sim(k_\cav\lm)^2$ and is thus of second order in the corresponding Lamb-Dicke parameter. In the following we will neglect this second order term and keep only the first one, where the cavity couples linearly to the membrane fluctuations.

The linear-coupling term for the atom vanishes, as \textit{for a single cavity mode} the atomic equilibrium position is defined by $\left[\partial_\at \omega_{\cav}\right] = 0$. Thus, only the quadratic, parametric term given in the last line of Eq.~\eqref{Eq:LinHam} contributes to the atom-cavity coupling. It is perfectly possible to proceed from here and to derive an effective coupling of the atom to the membrane. However, this coupling will be $\sim\Xm\Xa^2$ and thus not of the desired form given in Eq.~\eqref{Eq:Hatmec}.

\paragraph*{Two Mode Setup:} Creating a linear atom-cavity coupling requires non-vanishing cavity field slopes at the mean position of the atom, $\left[\partial_\at \omega_{\cav,i}\right] \neq 0$. To this end one has to require an external trap for the atom shifting it away from a lattice extremum. An elegant alternative is to use \textit{two} driven cavity modes ($i=1,2$) with the atomic equilibrium position at an extremum of the {\it combined} optical potential, and at the same time at a point of maximal slope of the individual cavity fields. Let us therefore study the cavity Hamiltonian $H_\cav$ in detail for this case of two driven cavity modes. In a frame rotating with the laser frequencies $\omega_i$ we have
\begin{multline}
H_\cav =
\sum_{i=1,2} \left[ \omega_{\cav,i} \left( \Xa,\Xm \right) - \omega_i \right]
A_i^\dag A_i  \\
+
\sum_i E_i \left( e^{i\phi_i} A_i^\dag + {\rm h.c.}\right),
\label{Eq:TWH}
\end{multline}
with mode frequencies $\omega_{\cav,i} \left( \Xa,\Xm \right)$ given by,
\be
\omega_{\cav,i} \left( \Xa,\Xm \right)
= \omega_{\cav,i}^0  - [g_{0,i}/\lm] \Xm + \U  u_i (\Xa).
\label{Eq:Hcavint}
\ee
The second term in (\ref{Eq:Hcavint}) describes the dynamic cavity detuning due to vibrational fluctuations of the thin dielectric membrane, with
single-photon coupling \cite{ThompsonNature2008,Jayich2008}
\be
g_{0,i} = f_i \left(\lm / L\right) \omega_{\cav,i}^0 .
\label{Eq:g0}
\ee
Here $L$ is the cavity length and $f_i=2 r \sin(2 k_i \Xmbar ) / [1 - r^2 \cos^2(2 k_i \Xmbar )]^{1/2}$ is a correction factor which takes into account the finite amplitude reflectivity $r$ of the membrane, as well as the distance $\Xmbar$ to the cavity field node where the field is zero and thus insensitive to the membrane motion.  Note that the special case $f_i=1$ is familiar from optomechanics with a perfectly reflecting moving mirror. By a proper choice of membrane location $\Xmbar$ it is possible to achieve $f_i \simeq 2r$ for both fields.

The third term in (\ref{Eq:Hcavint}) describes how the driven optical modes provide a lattice potential for the atom along the cavity axis, with the spatial intensity profile
\[
u_i(x) = \sin^2(k_i x),
\]
and a lattice potential strength determined by the AC Stark shift (per photon),
\be
\U = \Omega_0^2/\delta .
\label{Eq:U0}
\ee
Here $\Omega_0$ is the vacuum Rabi frequency and $\delta$ is the detuning of the lasers from the respective atomic transitions, assumed equal for simplicity (see Fig.~\ref{figAtomInt}).

The linearization of the dynamics around the equilibrium mean values is done as for the single-mode case discussed previously. The intracavity amplitudes are
$\alpha_i = E_i e^{i\phi_i}/(\di + i \kappa_i)$ with $E_i$ the drive strength of mode $i$, and the phase $\phi_i$ is chosen to make $\alpha_i$ real. $\Delta_{i}=\omega_i-\omega_{\cav,i}^0$ is the laser detuning relative to the cavity mode. The following derivation is in principle general regarding the number of driven modes, the mode parameters $\alpha_i,\kappa_i$ and the optomechanical coupling $g_{0,i}$. Without loss of generality we will assume in the following a symmetric two-mode case with $\alpha_i=\alpha$, $\kappa_i=\kappa$ and $g_{0,i}=g_0$.

The expansion is again very similar to the setup for a single mode. The main differences concern the atomic degrees of freedom. The atomic mean position $\Xabar$ is determined by vanishing first derivative of the {\it total} field,
\be
u^\prime (\Xabar)  =0 , \qquad u(x) = u_1 (x) + u_2 (x) .
\label{Eq:CondXabar}
\ee
and the trap frequency $\oma$ of the harmonic potential at this position is accordingly
\[
m \oma^2 = \U \alpha^2 k_1^2 \zeta (\Xabar) ,
\qquad  \zeta(x) =  u''(x)/ k_1^2.
\]
Most notably the ac Stark shift term $U_0\sum_i u_i (\Xa)A^\dag A$ gives rise also to a linear atom-cavity interaction, as the individual terms $\sim u_i^\prime (\Xabar)$ can be nonzero despite the condition on vanishing derivative of the total field (\ref{Eq:CondXabar}),
\be
u_{1(2)}\left( \Xa \right) \simeq
u_{1(2)}(\Xabar) \pm \eta \theta (\Xabar) (\aat + \aad)
\ee
with $\theta (x) =  u_1'(x)/k_1$ and Lamb-Dicke parameter $\eta=k_1\la$. A significant slope $\theta(\Xabar)$ and hence significant coupling is achieved for two modes by a careful choice of atomic site within the cavity, far from extremum points of the individual lattice modes, as we discuss in Sec.~\ref{Sec:Techdet}.

In this way it is straightforward to expand and linearize the cavity Hamiltonian  for the two mode setup in Eq.~\eqref{Eq:TWH}. When combined with the kinetic energy of the atom this results overall in a linearized Hamiltonian $H_{\rm sys}$ of the full master equation in Eq.~\eqref{Eq:MEW}
\be
H_{\rm sys} = H_0 + H_{\rm int}
\label{Eq:Hlinear}
\ee
with a free energy
\[
H_0= -\sum_i \di \aid \ai+\omm\amd\am+\oma\aad\aat,
\]
and \textit{linear} membrane-cavity \textit{and} atom-cavity interaction
\begin{align*}
H_{\rm int}&=\Gm(\am + \amd)[(\aone + \aoned)+(\atwo + \atwod)]\\
&\quad+\Ga(\aat + \aad)[(\aone + \aoned)-(\atwo + \atwod)]
\end{align*}
with coupling strengths,
\begin{align}
\Ga &= \U \alpha \eta \theta (\Xabar), &
\Gm &= g_0\alpha.
\label{Eq:gai}
\end{align}
In $H_{\rm sys}$ all parametric coupling terms have been neglected, as they will be smaller by the atomic Lamb-Dicke factor $\eta$ or by the much smaller Lamb-Dicke factor corresponding to the membrane motion, as discussed previously. In the limit of large cavity amplitude we also drop all terms of zeroth order in $\alpha$. For later use it will be convenient to reexpress the interaction in the form
\[
 H_{\rm int} =  g  \sum_i  \left[ \Fi + \Fid \right](\ai + \aid) ,
\]
with operators $\Fi$ describing the forces exerted by the atom and membrane motion on the cavity fields,
\begin{align}\label{eq:Force}
F_{1,2} = \left(- {\Gm \over g}\am  \pm {\Ga \over g}\aat \right) , \qquad g = \sqrt{\Gm^2 + \Ga^2}.
\end{align}

Before we derive the cavity-mediated atom-membrane coupling, we will finally discuss the atomic Lindblad term of Eq.~\eqref{Eq:LatFull}. In a Lamb-Dicke expansion around the atomic equilibrium position in the optical potential this Lindblad term takes the form of a momentum diffusion master equation
\[
L_\at (W) = {\ga \over 2} {\cal D}[\aat + \aad] (W),
\]
with a diffusion rate,
\be
\ga = \eta^2 s_e \gamma \left[ 2 - (4/5)u(\Xabar)\right]
\label{Eq:GammaAt}
\ee
where the saturation parameter is now explicitly given by $s_e = [\alpha \Omega_0 / \delta]^2$. The expression (\ref{Eq:GammaAt}) in the end depends on the particular atomic transition and the specific geometry; the factor $(4/5)$ is specific for transitions with $\Delta m = 0$ but for other transitions it is still of order unity.

Let us remark that in fact it is possible to solve the master equation of the full system exactly (e.g. by means of the methods given in Appendix~\ref{sec:timeevol}), and indeed there can be rich physics to be explored in the regimes not considered here. However, the focus of this paper is the regime where the cavity modes can be eliminated, $g_\at,g_\mec \ll {\rm max}\{\kappa,\Delta\}$, which is not only more relevant from an experimental point of view, but also allows for analytical, transparent results which highlight the physical properties of the system.

\subsubsection{Adiabatic Elimination of Cavity Field and Effective Master Equation}
\label{ssec:amint}

We are now in the position to derive an effective coupling mediated by the cavity modes. The idea is to use a parameter regime where the cavity dynamics is essentially unperturbed by the motion of the membrane and the atom, and solely mediates interaction between the two. The corresponding requirement is fast cavity dynamics, $g \ll \kappa$ {\it or} $g \ll |\di \pm \omm|$. For optomechanical cooling the former condition is the more common requirement, but since the resulting strong dissipation through the cavity decay would harm the coherent cavity-mediated dynamics, we choose a regime where $\di$ are the large parameters. Here fluctuations in the cavity quadratures are fast variables which adiabatically follow the dynamics of the position fluctuations of the atom and the membrane. In order to achieve strong interaction we further assume atom and membrane to be on resonance,
\begin{align}
\oma = \omm .
\label{Eq:ommoma}
\end{align}
The formal procedure for eliminating the optical modes, as described in detail in Appendix~\ref{App:Elim}, is to
perform adiabatic elimination
using standard techniques \cite{QuantumNoise}. We find that the
 linearized atom-membrane-cavity dynamics (\ref{Eq:Hlinear})
gives rise to the effective master equation (\ref{Eq:MEeff})
with
\[
H=\omm\amd\am+\oma\aad\aat  + H_{\at-\mec} .
\]
The last term $H_{\at-\mec}$ (\ref{Eq:HCmed}) represents the cavity-induced atom-membrane coupling and a correction to the free motion, and can be extracted from the coherent part of the cavity-mediated Liouvillian $L_{\rm c-med}$ (\ref{Eq:Lcmed}). In detail it reads,
\begin{multline}
H_{\at-\mec} = {i \over 2} \sum_i \Bigg[
{g^2 \over \kappa + i(\di - \omm)} \Fi \left( \Fi + \Fid \right) \\
+ {g^2 \over \kappa + i(\di + \omm)} \Fid \left( \Fi + \Fid \right)
- \hc\Bigg].
\label{Eq:Hngnting}
\end{multline}

The cavity decay translates into {\it correlated} decay $L_\cav (\rho)$ (\ref{Eq:DissCmed}) for atom and membrane, where in the rotating wave approximation (RWA) each optical mode $i$ contributes cooling (${\cal D}[\Fi]$) and heating (${\cal D}[\Fid]$) associated
with emission of sideband photons at either side of the driving
laser, that is, at one of the two frequencies $\omega_i \pm\omm$,
\begin{multline}\label{Eq:LcRWA}
 L_\cav(\rho) \simeq \sum_i \Bigg[ \left({g^2 \kappa \over \kappa^2 + (\di + \omm)^2 }\right) {\cal D}[\Fi](\rho) \\
+ \left({g^2 \kappa \over \kappa^2 + (\di - \omm)^2 }\right) {\cal D}[\Fid](\rho) \Bigg].
\end{multline}
An emission event is accompanied by the creation or annihilation of
a quantum in either the atomic motion or the membrane vibration. For a near resonant
system ($\omm \simeq \oma$) these two possibilities are indistinguishable,
such that both processes happen in a coherent fashion. Therefore,
the jump operators $\Fi$ are linear combinations of the corresponding
annihilation operators $\aat$ and $\am$.

\subsubsection{Coupling in the Dispersive Regime}
\label{sec:strongcoupl}

So far, we have derived expressions for the cavity-mediated interaction (\ref{Eq:Hngnting}) as well as its inevitable companion, dissipation through the cavity decay (\ref{Eq:LcRWA}). The remaining challenge is to reach the strong coupling regime for the reduced system, with effective coupling strength $\Geff$ which is much larger than all decay rates, $\Geff \gg \Gamma_\cav, \Gamma_\mec, \Gamma_\at$. Let us first consider the relation to the cavity-induced dissipation described by $L_\cav$.

Our first observation is that the atom-membrane coupling (\ref{Eq:Hngnting}) is maximized for equal and opposite detunings,
\[
\done = - \dtwo \equiv \Delta ,
\]
for which the two cavity modes respond equally and oppositely to the membrane vibration.
Evaluating the effective Hamiltonian (\ref{Eq:Hngnting}) for this special case we find
\begin{multline*}
H_{\at-\mec}  = - G \Big[(\am + \amd)(\aat + \aad) \\
 + i \varepsilon \left(\am\aat - \amd\aad\right)
\Big],
\end{multline*}
where we dropped a global energy shift. The effective coupling strength $G$ is given by
\[
G = \left[ { 2 \Gm\Ga (\Delta - \omm) \over \kappa^2 + (\Delta - \omm)^2} + { 2\Gm\Ga ( \Delta + \omm ) \over \kappa^2 + (\Delta + \omm)^2}\right].
\]

From the observation that the rate of cavity induced decoherence in $L_\cav$, see Eq.~\eqref{Eq:LcRWA}, scales like $\sim 1/\Delta^2$ whereas the cavity mediated interaction $\Geff \sim 1/\Delta$, we draw the conclusion that the dispersive limit is natural for suppressing dissipation. Focusing on the regime where $|\Delta|$ is the largest parameter, $|\Delta | \gg \omm , \kappa $, the correction $\varepsilon$ to a pure $(\am + \amd)(\aat + \aad)$-interaction is negligible,
\[
\varepsilon  =  {2 \kappa \omm \over  \Delta^2 + \kappa^2 - \omm^2}  \ll 1 .
\]
Thus the coherent dynamics in the reduced master equation \eqref{Eq:MEeff} is effectively given by the Hamiltonian $H$ in (\ref{Eq:Hatmec}).
This is the main result of our investigation. To zeroth order in $\kappa/\Delta,\omm/\Delta$ the coupling constant $\Geff$ has the simple form,
\[
\Geff \simeq {4 \Gm\Ga \over \Delta} .
\]
Regarding the cavity-induced decoherence processes described by Eq.~\eqref{Eq:LcRWA}, the combination of a red-detuned $(\done = \Delta < 0)$ and a blue-detuned $(\dtwo = - \Delta)$ laser drive can be interpreted as simultaneous cooling and heating processes. The rate of cooling $\Gamma_\cav^+$ via mode 1 equals the rate of heating via mode 2, and vice versa with rate $\Gamma_\cav^-$,
\begin{multline}
L_\cav (\rho)  = {\Gamma_\cav^+ \over 2}
\left( {\cal D}[\Fone](\rho) + {\cal D}[\Ftwod](\rho) \right)\\
+ {\Gamma_\cav^- \over 2}\left( {\cal D}[\Foned](\rho) + {\cal D}[\Ftwo](\rho) \right) ,
\label{Eq:L1L2}
\end{multline}
with the rates given by,
\[
\Gamma_\cav^\pm = {2\kappa \left(\Gm^2 + \Ga^2 \right) \over \kappa^2 + (\Delta \pm \omm)^2 } .
\]

In our attempt to minimize dissipation we additionally note that the relation between the coupling constants $\Ga$ and $\Gm$ is of importance. The ratio of dissipation to coupling strength is proportional to,
\[
\Gamma_\cav^\pm / \Geff \propto {\Gm^2 + \Ga^2 \over \Gm \Ga} .
\]
This implies that the mediated atom-membrane interaction is most efficient when the two oscillators couple equally strongly to the cavity modes, $\Gm = \Ga$. Under this condition and to lowest order in $\kappa/\Delta,\omm/\Delta$ the cooling / heating rates are in fact equal,
\be
\Gamma_\cav^\pm \simeq \ \Gamma_\cav = \Geff {\kappa \over \Delta} ,
\label{Eq:GammaPM}
\ee
a factor $\kappa/\Delta \ll 1$ smaller than the coupling constant $\Geff$.

\subsection{Alternative setups}

In this section we extend the previous discussion to give a hint about alternative mode configurations for the proposed setup. In particular we discuss how to obtain a \textit{time-dependent atom-membrane coupling} $\Geff(t)$, which can be advantageous e.g. for entanglement creation as discussed in section \ref{sec:cohevol}.

\subsubsection{Single driven mode combined with external trap}

As briefly mentioned previously, as an alternative to the two-mode setup one could use a single driven mode combined with an external atom trap. This trap would shift the atom away from the lattice extremum, to an equilibrium point $\Xabar$ where the cavity field has finite slope, $u_1' (\Xabar) \neq 0$, which can be significant for a properly chosen atom location within the cavity. With coupling only to a single mode, the force $\hat{F}_1$ is given by (\ref{eq:Force}) whereas $\hat{F}_2 = 0$. Consequently, with only one cavity mode mediating the membrane vibration, the largest displacement of the atomic mean position is only half as large compared to the case when two cavity modes are shifted out of phase. Hence the resulting coupling constant is only half of its maximum value,
\[
\Geff_{\rm single - mode} \simeq {2\Gm\Ga \over \Delta} .
\]

\subsubsection{Time-dependent coupling constant $\Geff(t)$}\label{Sec:TimeDepG}

In principle, it is possible to achieve a time-dependent coupling constant $\Geff(t)$ by modulating the laser power, resulting in a time-dependent intracavity amplitude $\alpha(t)$. However, the basic problem with simply introducing time-dependent $\alpha (t)$ in the two-mode setup, is that the atom potential will be time-dependent as well, which could heat up the atomic motion. Therefore we consider the following modified setup, where we either use an external trap for the atom as above, or use a second mode mainly to provide the trapping potential. Either way, the role of the first mode is to mediate the atom-membrane interaction with modulated strength, $\alpha_1 (t) = \alpha_1 c(t)$.

Without going into details, the idea of the two-mode case is to drive the second mode
such that the corresponding intra-cavity field becomes very strong, $\alpha_2 \gg \alpha_1$, with $\alpha_1$ the amplitude of the coupling mode. Due to the very small ratio $\alpha_1/\alpha_2 \ll 1$, the first mode hardly influences the atomic potential at all, and the mean atom position $\Xabar$ is given by $u_2^\prime (\Xabar)= 0$. Furthermore the atomic frequency is determined by the curvature of the second field, $m\oma^2 = \alpha_2^2 u_2''(\Xabar)$. Since $u_2^\prime = 0$ at the atomic equilibrium point, the second mode will not contribute to the linear atom-cavity coupling.

With the atom-membrane interaction mediated by only a single mode, as discussed above, the coupling constant will be half of the maximum value. We find the resulting time-dependent atom-membrane coupling,
\[
\Geff_{\rm two-mode} (t) \simeq {2\Gm\Ga \over \Delta}c^2(t).
\]
We will come back to this possibility of making the coupling explicitly time dependent in our discussion of coherent evolution, in particular of a protocol to generate entangled states of the atom and the membrane, see Sec.~\ref{subsec:entang}.

We will remark on yet another type of setup in the outlook of this paper (Sec.~\ref{Sec:Disc}), namely how to implement an optomechanical \textit{Jaynes-Cummings model} where atomic \textit{internal} degrees of freedom are used instead of the atomic motion. Together with the examples of this section, the discussion illustrates that the present setup actually provides a toolbox for engineering various interactions and different types of dynamics.

\section{Coherent evolution in the strong coupling regime}
\label{sec:cohevol}

In the previous section it was shown that we can implement a linear atom-membrane interaction (\ref{Eq:Hatmec}) with the proposed two-mode setup operated in the dispersive regime, and that this interaction can be fast on the time scale of relevant decoherence rates in this system. In this section we will study a few applications, which become accessible in this regime.

Note first that the coherent evolution governed by this Hamiltonian transfers a state from the atom to the membrane, and vice versa, in a time $\ts$ given by
\[
\ts = \pi/(2 \Geff),
\]
such that $|\psi_1\rangle_\at |\psi_2\rangle_\mec \rightarrow |\psi_2\rangle_\at |\psi_1\rangle_\mec$, up to local rotations. The state swap mechanism appears naturally in the interaction picture; for resonant coupling $\oma = \omm \gg \Geff$ the Hamiltonian takes a beam-splitter form in the RWA,
\[
H_{\rm I} \simeq \Geff \left( \aat \amd + \hc \right) .
\]
Particularly intriguing is the ability to use the state transfer to control the mechanical state through the available atomic physics toolbox.
However, as already discussed, the coherent interaction is accompanied by several sources of noise which in the end reduce the fidelity of the state transfer. Strong coupling is therefore established by fulfilling, additionally to the resonance condition, the following set of conditions,
\begin{align}
\Geff \gg \Gamma_\at,\Gamma_\mec,\Gamma_\cav .
\label{Eq:Cond0}
\end{align}
In section \ref{ssec:optim} we will summarize and comment on the optimization of parameters which is necessary in order to reach the strong coupling regime, following \cite{hammerer_strong_2009}.
Here we illustrate the strong coupling in the presence of noise with three specific examples of state transfer from atom to membrane: coherent and squeezed state as well as a Fock state. Aiming at a clear picture of the effect of dissipation, we assume all dissipation rates to be equally strong  and define a ratio $f$,
\be
f = {\Gamma_\cav \over \Geff} = {\ga \over \Geff} = {\Gamma_\mec \over \Geff }.
\label{Eq:eqdiss}
\ee
Note that we here put the membrane heating and cooling rates equal, $(\nm+1)\gm \simeq \nm\gm = \Gamma_\mec $,
and assume the dispersive regime where $\Gamma_\cav^\pm \rightarrow \Gamma_\cav$. Our aim is to find the acceptable noise level which still allows for state transfer, by solving the master equation (\ref{Eq:MEeff}) with the Hamiltonian (\ref{Eq:Hatmec}) exactly. These numerical solutions are combined with analytical calculations based on the RWA as described above. In fact, for Gaussian states it is straightforward to analytically solve for the time evolution; the details of the derivation are presented in Appendix \ref{sec:timeevol}. Interestingly, with the generalized technique presented in Appendix \ref{subsec:Fock}, the impact of noise on the evolution of Non-Gaussian states, e.g. Fock states can be derived as well. In all three examples the noise introduces a thermal population $\bar{n}_{\rm s}$ during the time-interval $\ts$ needed for state transfer. We find,
\[
\bar{n}_{\rm s} = \pi f .
\]
Another interesting application is cooling of the membrane via coherent state swap, as will be presented in subsection \ref{subsec:cool}. Finally in \ref{subsec:entang} we present a way to entangle atom and membrane, using a time-dependent coupling $\Geff(t)$ -- as discussed in Sec.~\ref{Sec:TimeDepG} -- which enhances exactly those terms which are neglected in the RWA.

\subsection{Coherent state swap}

\begin{figure}[ht]
\begin{centering}
\includegraphics[width=\columnwidth]{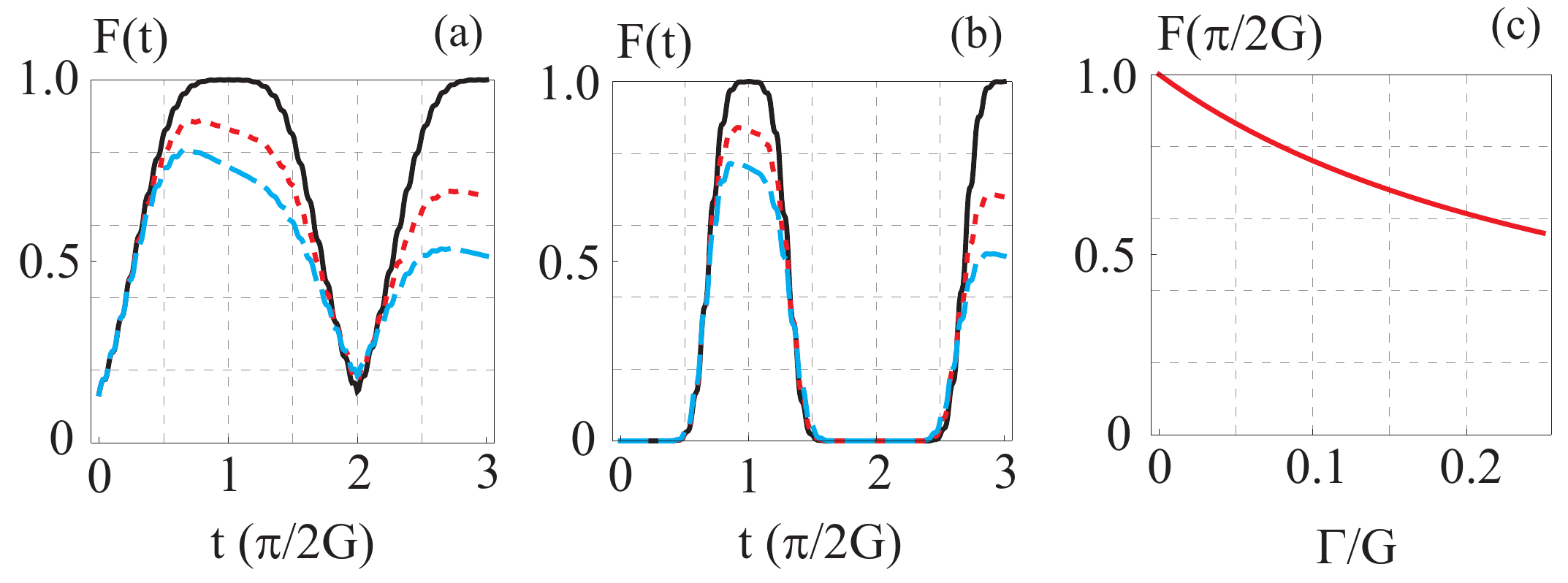}
\caption{Fidelity for transfer of coherent state $|\beta\rangle$ from atom to membrane.
(a,b) Fidelity as function of time for transfer of a state with (a) $\beta = 1$ and (b) $\beta = 5$ for various values of the dissipation ratio $f$, with fixed $\Geff/\omm = 0.034$. Here $f=0.01$ (black solid line), $f=0.05$ (orange dotted line) and $f=0.10$ (blue dashed line). The little wiggles are due to counter-rotating terms. (c) Snapshot at $t = \pi/2G$: fidelity for transfer of state with $\beta = 1$ as a function of the dissipation ratio $f$. }
\label{figCoh}
\end{centering}
\end{figure}

Our first example is the transfer of a coherent state $|\beta\rangle$ from the atom to the membrane. The perfect state swap evolves a state $|0\rangle_\mec |\beta\rangle_\at$ into $|\beta e^{i\phi(t)}\rangle_\mec |0\rangle_\at$ with the phase $\phi(t)$ governed by the system Hamiltonian. Here we use the fidelity $F$, defined as the overlap between the original atomic wave function and the final membrane wave function, as a figure of merit for the effect of noise during the state transfer. Figures \ref{figCoh}(a,b) show the fidelity for transfer of two different coherent states, $|\beta = 1\rangle$ and $|\beta = 5\rangle$. Particularly interesting is the fidelity of the state swap, i.e. for $t=\ts$ where $F$ is close to maximal but still deteriorated due to dissipation. In fact, the dependence of the state swap fidelity (at $t=\ts$) on the noise ratio $f$ follows the analytical result (\ref{Eq:fidgeneral}) derived in Appendix \ref{sec:timeevol} in the RWA,
\be
F \left(\ts \right) = {1 \over 1 + \pi f}.
\label{Eq:fid}
\ee
This simple analytical result, which very well matches the exact numerical solution shown in Fig. \ref{figCoh}(c), states that with noise levels below 10\% we can expect a state transfer fidelity above 75\%.

\subsection{Squeezed state transfer}

\begin{figure}[ht]
\begin{centering}
\includegraphics[width=\columnwidth]{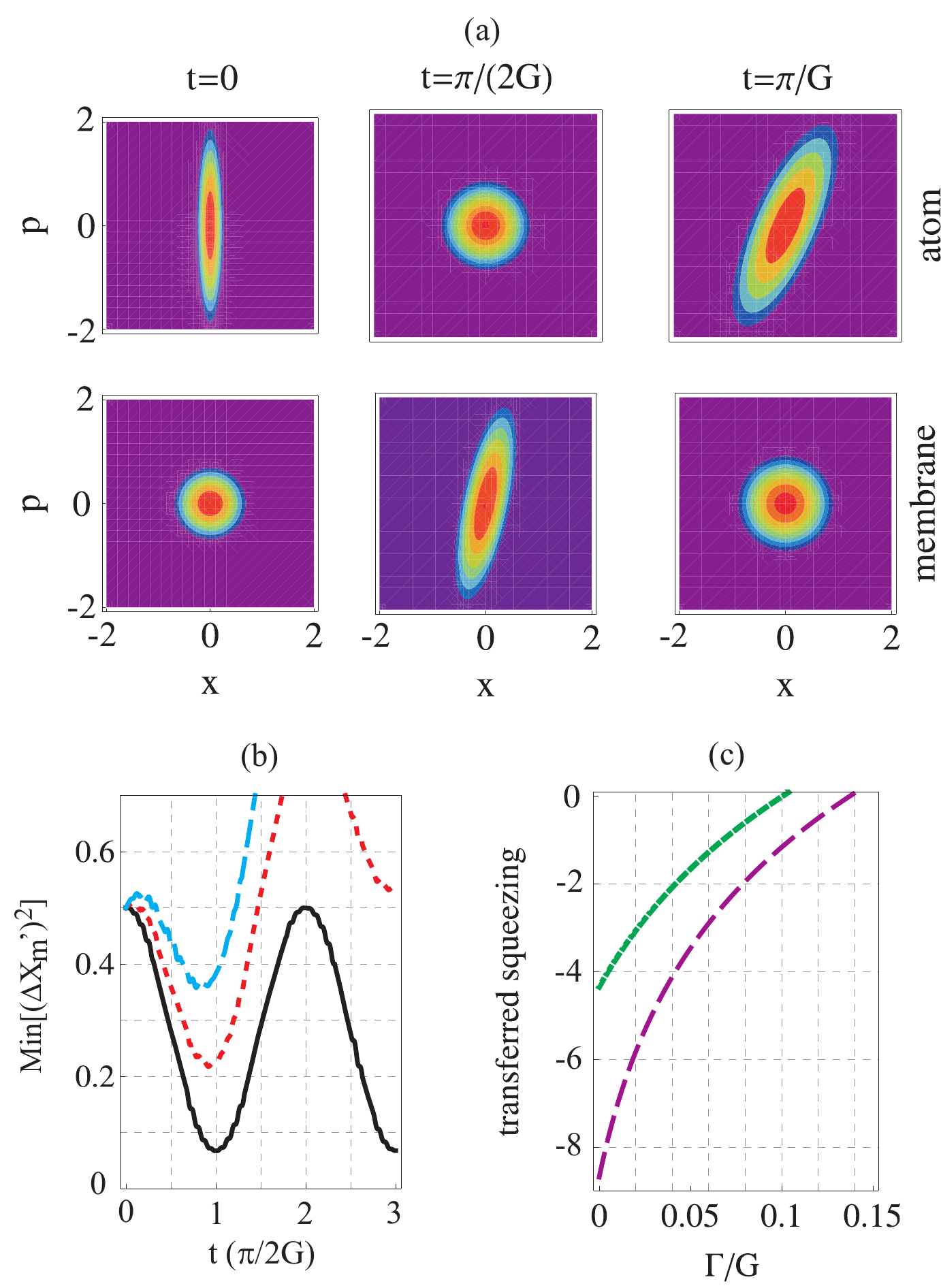}
\caption{Transfer of squeezed atom state with initial variance $\Delta X_\at^2=e^{-2}/2$ to the membrane. (a) Snapshot of Wigner functions (upper row - atom, lower row - membrane) for $f=0.05$ at $t=0$, $t=\pi/2G$ and $t=\pi/G$. (b) Minimum membrane variance as function of time for different dissipation ratios $f=0$ (black solid line), $f=0.05$ (orange dotted line) and $f=0.10$ (blue dashed line). (c) Transferred squeezing (in dB; $S(t) = 10 \log_{10}[s(t)]  ({\rm dB})$) as function of the dissipation ratio $f$, for fixed  $\Geff/\omm = 0.034$. The initial atom squeezing is given by $s(0) = e^{-2}$ (purple dashed line) corresponding to $-8.7$ dB, and $s(0) = e^{-1}$ (green solid line) corresponding to $-4.3$ dB, respectively.  }
\label{figSqueez}
\end{centering}
\end{figure}

The second example is the transfer of an atomic squeezed state $|\xi\rangle$ with minimal variance $\Delta X_\at^2 = (1/2)s$; here $s<1$ denotes a state squeezed along the $X$ quadrature. Such a state can be constructed using for example the parametric coupling to the cavity field, $\sim (\aat + \aad)^2 (\ai + \aid)$ which was briefly mentioned in section \ref{Sec:effMA}. Ideally the swap operation transfers the atomic minimal variance to the membrane state, $|\beta\rangle_\mec |\xi\rangle_\at \rightarrow |\xi e^{i\varphi}\rangle_\mec |\beta e^{i\phi}\rangle_\at$. Dissipation however broadens the variance during the swap operation,
\[
\Delta \left(X^\prime_\mec \right)^2 = {1 \over 2}s(\ts),
\]
with $s(\ts)> s(0)$. Fig. \ref{figSqueez}(a) shows the minimal variance of the membrane, reaching its lowest value after half a period ($t=\ts$) when the squeezed state has been transferred from the atom to the membrane. Obviously larger dissipation ratio $f$ results in less squeezing transferred.
In Appendix \ref{sec:timeevol} we derive the following analytical expression in the RWA for the dependence of the squeezing parameter $s$ on the dissipation rate $f$,
\[
s\left(\ts \right) = s(0)  + 2 \pi f  .
\]
Fig. \ref{figSqueez}(b) shows snapshots of the atom and membrane Wigner functions at $t=0$, $t=\ts$ and $t=2\ts$; one clearly sees how the dissipation broadens the variances. Furthermore, due to the coherent evolution, the squeezed membrane quadrature $X^\prime_\mec$ is not necessarily equal to the squeezed atom quadrature $X_\at$.

In Fig. \ref{figSqueez}(c) we show how the membrane minimal variance increases with the noise ratio $f$ for two specific examples of initial minimal variance of the atom.
The exact result confirms the loss of squeezing given by the expression above for $s(\ts)$. With noise levels below $f \sim 10$\% an initial atom squeezing below $-4.3$ dB allows for squeezing of the membrane.
%

\subsection{Fock state transfer}

The previous two sections dealt with the engineering of Gaussian states of the mechanical resonator, while the ultimate goal would of course be to apply these methods to create more non-classical state, e.g. states with negative Wigner functions.

\begin{figure}[ht]
\begin{centering}
\includegraphics[width=\columnwidth]{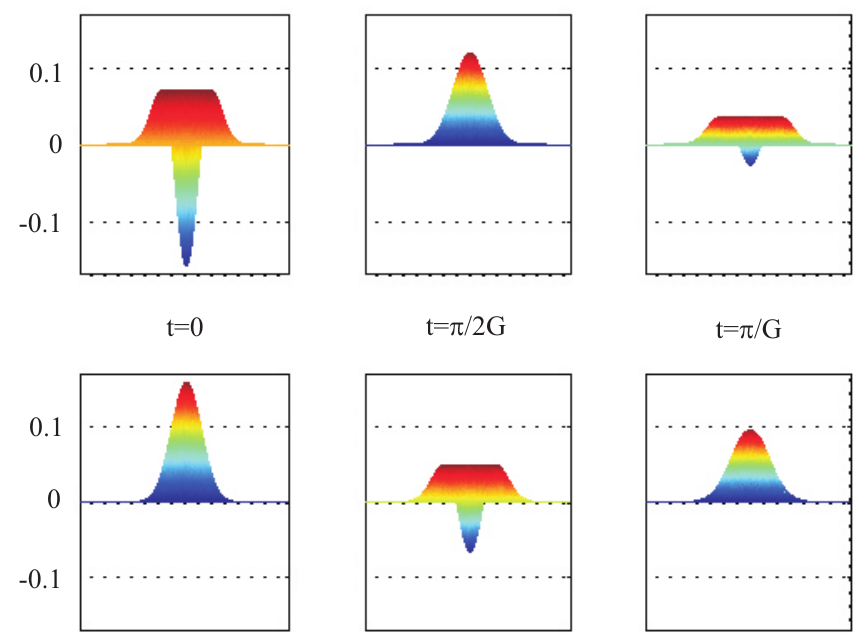}
\caption{Transfer of the Fock state $|n=1\rangle$ from atom to ground-state prepared membrane. Snapshot cuts through the Wigner functions at times $t=0$, $t=\pi/2G$ and $t=\pi/G$, for fixed dissipation ratio $f=0.05$ and $\Geff/\omm = 0.034$. }
\label{figFock}
\end{centering}
\end{figure}

\begin{figure}[ht]
\begin{centering}
\includegraphics[width=0.8\columnwidth]{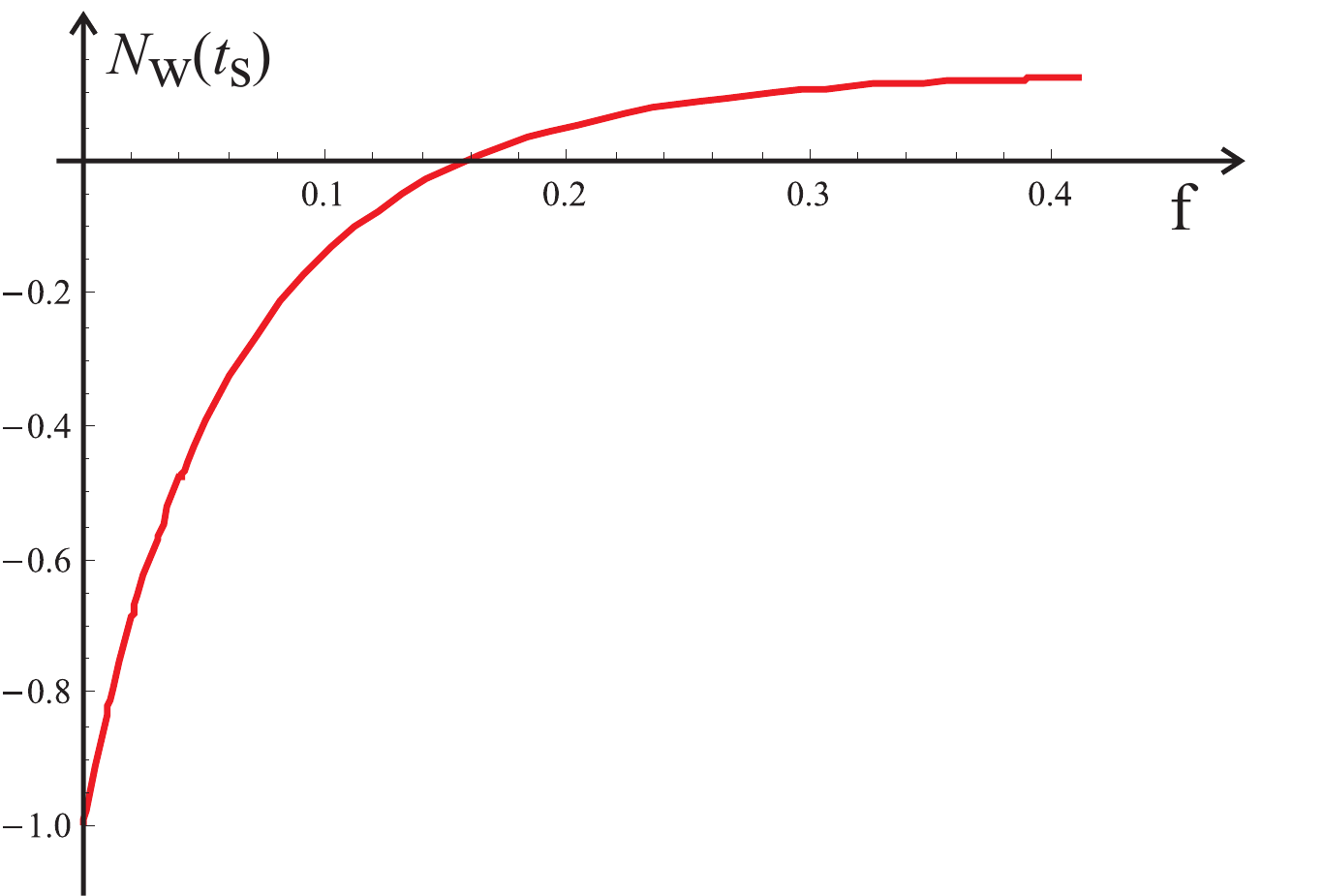}
\caption{Analytical result for the relative membrane Wigner function negativity $N_{\rm w} $ at time $t=\ts$ as a function of the dissipation ratio $f$.}
\label{figNeg}
\end{centering}
\end{figure}

As a last example we therefore present the transfer of a Fock state with $n=1$ from the atom to the membrane. Assuming the membrane to be ground-state cooled, the ideal evolution reads $|0\rangle_\mec |n=1\rangle_\at \rightarrow |n=1 \rangle_\mec |0\rangle_\at$. The quantum properties of the Fock state are best illustrated by the negative value of its Wigner function at the origin. Fig. \ref{figFock} shows cuts through the Wigner function for three instants in time, $t=0$, $t=\ts$ and $t= 2\ts$.
Thermalization during the state transfer is reflected in the decreasing Wigner function negativity for each state swap. A convenient figure of merit for the thermalization is therefore the value of the membrane Wigner function $w_\mec(\beta,\beta^*,t)$ at $\beta=\beta^* = 0$ relative to the corresponding (absolute) value for a Fock state $w_{\rm F} (0,0)$,
\[
N_{\rm w} (t) \equiv {w_\mec(0,0,t)  \over |w_{\rm F} (0,0)| }.
\]
An analytic expression for $N_{\rm w} (t)$ in the RWA is derived in Appendix \ref{sec:timeevol}, see Eq. (\ref{Eq:FockNeg}). In Fig.~\ref{figNeg} we present a case of particular interest, namely the membrane Wigner function negativity after the first state swap ($t=\ts$), which depends on the dissipation ratio $f$ according to,
\[
N_{\rm w} \left(\ts \right) =
{-1 + 2 \pi f
\over \left( 1 + 2 \pi f \right)^2}.
\]
The quantum properties of the Fock state $|n=1\rangle$ are transferred to the membrane if the dissipation ratio is sufficiently small, $2\pi f < 1$. For an experimentally feasible noise ratio $f=0.1$, 14\% of the Wigner function negativity is preserved during the swap operation.
%

\subsection{Membrane cooling through state swap}
\label{subsec:cool}

With the present setup, we see two routes towards preparing the membrane ground state. The first route is along the lines of cavity cooling; for example cooling the membrane via the cavity decay,
or via an externally controlled Raman atom cooling with rate $\Gamma_{\rm R}$ in the
ground state cooling regime $\Geff \ll \Gamma_{\rm R} \ll \omm$. The second route is to perform a state swap and hence transfer the ground state to the membrane from the (previously cooled) atom. Comparing the two routes, we find that the effective rate for state swap is much higher than for cooling, $ \Gamma_{\rm swap} \sim \Geff \gg \Gamma_{\rm cool} \sim \Geff^2/\Gamma_{\rm R}$, and that the state swap leads to a final occupation which is a factor $\Geff/\Gamma_{\rm R}$ lower than for cooling,
\[
\bar{n}_{\rm f}^{\rm swap} \sim \pi f + \left( {\Geff \over 2\omm}\right)^2 ,\qquad
\bar{n}_{\rm f}^{\rm cool} \sim f {\Gamma_{\rm R} \over \Geff} + \left( {\Gamma_{\rm R}\over 2\omm}\right)^2.
\]
Note that the expression for $\bar{n}_{\rm f}^{\rm swap }$ can be
optimized with respect to $\Geff$, since a large coupling strength on one hand decreases the noise level $f$ but on the other hand increases the residual final occupation in the second term. Replacing $f$ with  $\Gamma_\cav/\Geff$ and considering $\Gamma_\cav$ as a fixed parameter, one obtains the minimum $\bar{n}_{\rm f}^{\rm swap}$ for
$G_{\rm opt}=\sqrt[3]{2\pi \Gamma_\cav\omega_\mec^2}$.

Concluding that state swap cooling is more efficient than indirect Raman
cooling, it is still interesting to make a comparison with typical
cavity cooling in the good cavity regime where now $\kappa \ll \omega_\mec$
and in the perturbative regime where $\Gm \ll \kappa $. In this case, the
cooling rate scales as $\Gamma_\cav \sim \Gm^2/\kappa $ and the
final occupancy is
\[
\bar{n}_{\rm f}^\cav \sim \frac{\Gamma_\mec \kappa }{\Gm^2} + \left( \frac{\kappa }{2\omm}\right)^2.
\]
We conclude that $\bar{n}_{\rm f}^{\rm swap} < \bar{n}_{\rm f}^\cav$ for large enough atom-membrane coupling, $\Geff/\pi > \Gm^2/\kappa $.

One drawback with the state swap procedure is that it only works for a
precooled membrane; the anharmonicity of the atom well supports the transfer
of only a few quanta from the membrane, say $n_{\rm well}\sim 5-10$.
The situation looks better in a generalized setup with $N$ atoms distributed
over the lattice. In the ideal case of no atom-atom interaction and
identical atom site conditions, the effective coupling $\Geff$ is enhanced by
a factor $\sqrt{N}$ and the single atom operator $a_{\at}$ can be
substituted by the center-of-mass operator $A_{\rm cm }=(1/\sqrt{N})\sum_{j=1}^{N}a_{\at,j}$. Here we can consider introducing even further anharmonicity of the atomic wells to prevent an individual atom to
be multiply excited, in which case the center-of-mass mode can support the transfer and storage of $N$ excitations from the hot membrane, thus allowing the swap of a fairly
large thermal membrane occupation.

\subsection{Entanglement}
\label{subsec:entang}

In this subsection we lay out the prospects of observing entanglement between the atom and the membrane. The major obstacle in this regard is the coupling of the system to the environment. Even when we assume the membrane to be prepared in the ground-state initially, the system will quickly heat up and entanglement is lost, at
least if one just considers the usual static coupling.Here we point out, instead, a method for generating entanglement based on a time-dependent modulation of the input laser intensity that controls the atom-membrane interaction strength. In fact, this scheme can be employed generally in optomechanically coupled mechanical systems. The scheme turns out to be relatively robust against the impact of the dissipation channels. By modulating the atom-membrane coupling strength in time one can realize a non-degenerate parametric amplifier (two-mode squeezing) which induces strong quantum correlations between atom and membrane despite the simultaneously occuring heating of the system. We consider the linear membrane-atom interaction Eq. (\ref{Eq:Hatmec}) with the coupling constant modulated according to,
\[
\Geff(t)=\Geff\cos^2(\bar\omega t), \qquad \bar\omega = \frac{\omm +\oma}{2} .
\]
In order to allow for a modulation of the coupling strength without modulating the trapping frequency the setup needs to be modified as discussed in Sec.~\ref{Sec:effMA} B. Switching into the interaction picture, we find that in contrast to the case of constant coupling $\Geff$ previously discussed,
the coupling term here effectively transforms into a parametric amplifier part and a slowly oscillating beam splitter part (in RWA),
\[
H_{\rm I} \simeq
\frac{\Geff}{4}[\am \aat +{\rm h.c.}]
+ \frac{\Geff}{2}[\amd \aat e^{i(\omm-\oma)t}+{\rm h.c.}],
\]
and contributions that are oscillating fast with respect to the time-scale of $\Geff$,
and hence have negligible influence. It is the parametric amplifier
part that can be exploited to generate strong correlations.

As a measure of entanglement we employ the logarithmic negativity
\citep{1999_Eisert_EntanglementMeasures_JModOpt,2000_Simon,2002_Vidal_Entanglement}.
For a Gaussian state it can be computed directly from the elements
of the covariance matrix \citep{2002_Vidal_Entanglement}, whose time evolution was derived in Appendix B.
\begin{figure}[h]
\includegraphics[width=\columnwidth]{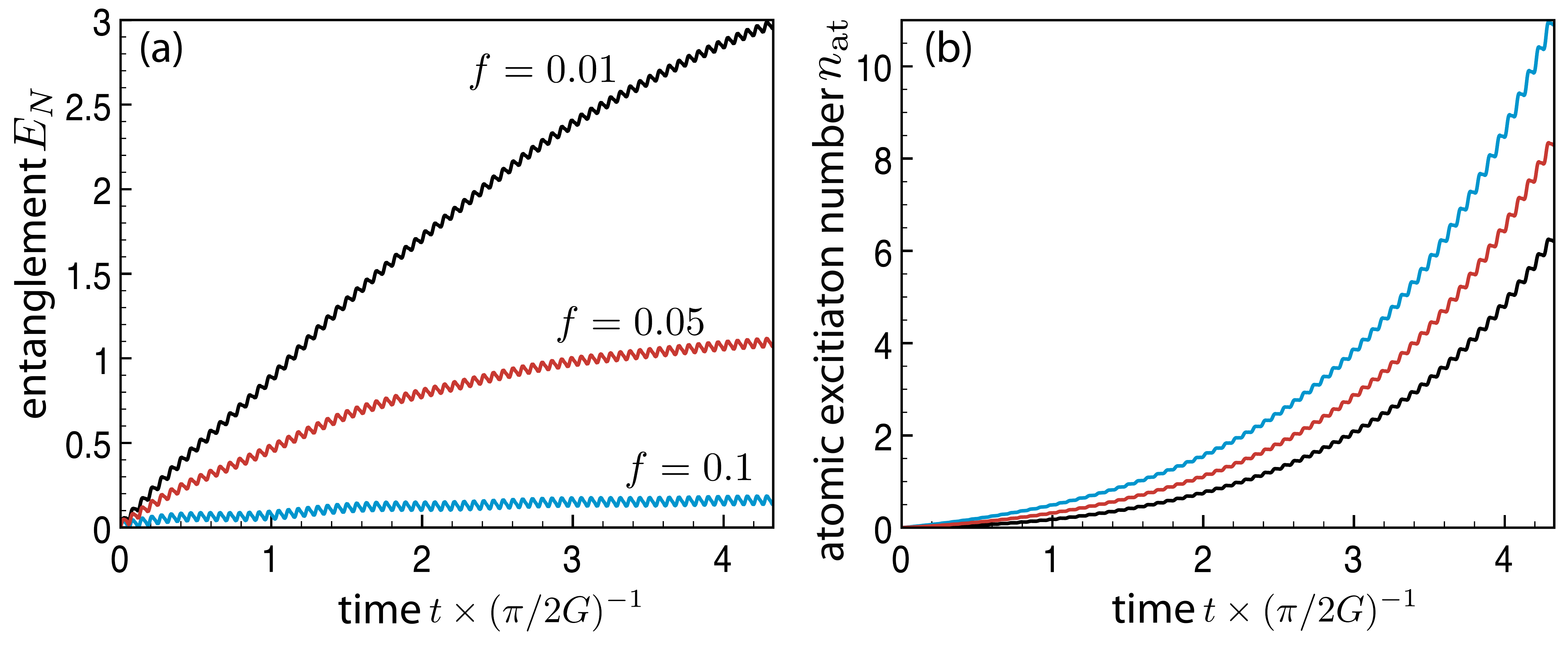}
\caption{Generation of entanglement between atom and membrane by modulating
the coupling strength. (a) Logarithmic negativity as a function of
time for different values of the dissipation ratio $f=0.1$ (blue curve), $f=0.05$ (red curve) and $f=0.01$ (black curve). (b) Corresponding increase of the
atomic excitation number $n_{{\rm at}}=\langle \aad \aat\rangle$.
For these plots we chose $\Geff/\omm=0.034$ and $\oma/\omm=1.1$
and assume the membrane to be in the ground state initially.}
\label{Flo:Entanglement}
\end{figure}

Fig. (\ref{Flo:Entanglement}) displays the generation of entanglement
(a) and the increase of the atomic excitation number $n_{{\rm at}}=\langle \aad \aat \rangle$
(b). This number should not exceed a threshold value of $n_{{\rm at}}\sim5-10$
in order to keep the effects of the anharmonicity of the trap negligible.
Note that we chose a relatively large difference in the oscillation frequencies, $\oma/\omm=1.1$,
in order to suppress the influence of the beam splitter interaction.
As indicated by this numerical example, driving the system with time-modulated driving strength
provides a useful method for generating entanglement in a quantum system that is in contact with a thermal bath.

For the parameters discussed here, the rates of the optomechanical cooling and heating processes are equal ($\Gamma_\cav^{\pm} \simeq \Gamma_\cav$), and these processes reduce the entanglement. We note, however, that the optomechanical damping can in principle also be used to generate entanglement in a steady-state situation (i.e. for fixed coupling $\Geff$) by reducing the effective occupation numbers of the mechanical oscillators as discussed in \citep{2009_EntanglementNoneqBath}.

\section{Technical details regarding the experimental setup}
\label{Sec:Techdet}

In this section we discuss technical issues regarding the specific setup that we have in mind, namely the optimal atom location within the cavity field, and the membrane heating due to power absorption.

\subsection{Optimization of parameters}
\label{ssec:optim}

Demanding strong atom-membrane coupling, i.e. fulfilling the conditions $\Geff \gg\Gamma_\cav,\Gamma_\at,\Gamma_\mec $ (\ref{Eq:Cond0}), in the end boils down to satisfying constraints on the cavity and membrane geometry, choosing the proper detunings and finding suitable atomic transitions. In the following we will therefore go through the set of conditions $\Geff \gg\Gamma_\cav,\Gamma_\at,\Gamma_\mec $ and $\oma=\omm$ (\ref{Eq:ommoma}) in detail.

In order to obtain weak cavity-induced decay $\Gamma_\cav \ll \Geff$, we concluded in Sec. \ref{Sec:effMA} that it is necessary to drive the cavity far off resonance
\begin{align}
\Delta \gg \kappa,\omm,
\label{Eq:Cond1}
\end{align}
and to keep at the same time a balanced atom--cavity and membrane--cavity
coupling $\Gm \simeq \Ga$, which is equivalent to demanding
\[
g_0 \alpha \simeq \U \alpha \eta \theta(\Xabar) .
\]
Here we first note that the intracavity amplitude $\alpha $ drops out. For simplicity we estimate $f_i \simeq 2 r$ and $\theta(\Xabar) \simeq 1$. In the following we insert the respective definitions of $g_0$ (\ref{Eq:g0}) and $\U$ (\ref{Eq:U0}) and use that $\omega_\cav \simeq ck_1$, and find
\[
2 r {c \over \kappa L} \lm \simeq {\Omega_0^2 \over \kappa\delta }\la .
\]
The difference in zero-point fluctuations between membrane and atom will give a factor $\lm/\la = \sqrt{m/M}$. Moreover, on the left side, we introduce the cavity finesse ${\cal F}$,
\[
{\cal F} = {\pi c \over 2\kappa L}
\]
and on the right side the cooperativity parameter,
\[
C = {\Omega_0^2 \over \kappa \gamma}.
\]
This way, the ratio of the coupling constants $\Gm/\Ga$ turns into a ratio of the cavity finesse ${\cal F}$ to the reduced single-atom cooperativity $\left( \gamma/\delta\right)C$, which must be balanced by the mass ratio $m/M$,
\begin{align}
\left( \frac{4r}{\pi}\mathcal{F}/  {\gamma \over \delta}C \right)\sqrt{\frac{m}{M}}\simeq1.
\label{Eq:Cond2}
\end{align}
The equality (\ref{Eq:Cond2}) does not only put a condition for weak cavity-induced decay, but will also be useful in the following to connect the respective parameters related to the membrane-cavity and atom-cavity coupling.

Here we should comment on the dependence of the ratio ${\cal F}/C$ on the cavity geometry. At first glance it may seem like this ratio can be controlled through the cavity length $L$, with ${\cal F}/C \sim (c/L)(\gamma/\Omega_0^2)$. This is however not the case. Keeping in mind that the electric field strength is proportional to $\sqrt{1/V}$, with the mode volume $V= A L $ and $A$ the beam cross section, the dependence on the cavity length $L$ in the cooperativity $C$ through the relation $\Omega_0^2 \sim 1/(A L)$ in fact cancels the length dependence in ${\cal F}$, assuming fixed cross section $A$. We find that the relevant geometric parameter for this ratio is the beam cross section,
\[
{{\cal F} \over C} \sim A .
\]

Next we require small decoherence due to atomic momentum diffusion, $\Gamma_{\at}/\Geff \ll 1$, which gives the condition,
\be
{4\Ga\Gm \over \Delta} \ll \eta^2 p_e \gamma .
\label{Eq:condat}
\ee
First of all, we use the condition on the coupling constants $\Ga \simeq \Gm$ to write the inequality (\ref{Eq:condat}) in terms of atomic parameters,
\[
{4\left(\U \alpha \eta \right)^2\over \Delta} \ll \eta^2 p_e \gamma .
\]
The Lamb-Dicke parameter $\eta$ drops out. Furthermore, with $p_e = \alpha^2 \U/\delta$, the intracavity field amplitude $\alpha$ also drops out, and what remains is a condition on the cooperativity parameter,
\begin{align}
C\gg\frac{\Delta}{4\kappa},
\label{Eq:Cond3}
\end{align}
which has to be very large, taking into account the condition (\ref{Eq:Cond1}).

Finally, thermal decoherence depends on the ambient temperature $T$
of the membrane. As we will discuss in more detail further below (see Sec.~\ref{subsec:absorp}), it is reasonable to assume that heating of the membrane is in fact caused dominantly by absorption of laser power, which depends in particular on the thermal link $\kappa_{\rm th}$ of the membrane to its support. The condition of small thermal decoherence $\Gamma_\mec/\Geff \ll 1$ reads in detail,
\be
{4\Ga\Gm \over \Delta} \gg {k_B T \over \hbar Q_\mec} = {\gamma_\mec \over \omm}{2\pi \over \kappa_{\rm th}{\cal F}}{\omega_\cav c \alpha^2 \over L} .
\label{Eq:condmem}
\ee
First of all, we use the condition on the coupling constants $\Ga \simeq \Gm$ to write the inequality (\ref{Eq:condmem}) without atomic parameters. Using also that $\omm = \hbar/(2M\lm^2)$, we arrive at,
\[
{(2r)^2 \lm^2 \omega_c^2 \alpha^2 \over  L^2 \Delta} \gg { M \lm^2 \gamma_\mec \over \hbar^2 }{\pi \over \kappa_{\rm th}{\cal F}}{\hbar\omega_\cav c \alpha^2 \over L} .
\]
The amplitude $\alpha$ and the zero-point fluctuation $\lm$ drop out of the inequality. In the same fashion as for the condition (\ref{Eq:Cond3}), we rewrite the inequality with respect to $\Delta/\kappa$,
\begin{align}
\frac{8r^{2}\mathcal{F}^{2}}{\pi^{2}}\frac{\kappa_{th}}{\gamma_{m}}\frac{\hbar\omega_{c}}{Mc^{2}}
\gg\frac{\Delta}{\kappa},
\label{Eq:Cond4}
\end{align}
with the first factor related to the properties of the membrane in the cavity, the second factor comparing the thermal link of the membrane to its natural linewidth, and the third factor comparing the energy of a single cavity photon to the effective ``rest energy'' of the membrane. Remarkably, this condition is independent of the laser power, and the left-hand side depends only on parameters fixed at fabrication.

Together, Eqs.~(\ref{Eq:Cond1}), (\ref{Eq:Cond2}), (\ref{Eq:Cond3}) and (\ref{Eq:Cond4}) ensure the set of conditions for strong coupling in \eqref{Eq:Cond0}. Note that the intracavity amplitude $\alpha$ dropped out in all cases. The absolute timescale of the system is thus not fixed by Eqs.~(\ref{Eq:Cond1}), (\ref{Eq:Cond2}), (\ref{Eq:Cond3}) and (\ref{Eq:Cond4}), but by the resonance condition $\omm=\oma$ which fixes the cavity amplitude $\alpha$,
\[
\omm \simeq \eta^2 \alpha^2 {\Omega_0^2 \over \delta} .
\]
The membrane frequency $\omm$ is fixed by construction, whereas the intracavity amplitude depends on the laser power $P$,
\[
\alpha^2 \simeq \left( {\kappa \over \Delta}\right)^2 {2P \over \kappa \hbar\omega_\cav} .
\]


\subsection{Details of ac Stark shift potential}
\label{subsec:Starkpot}

\begin{figure}[h]
\begin{centering}
\includegraphics[width=0.9\columnwidth]{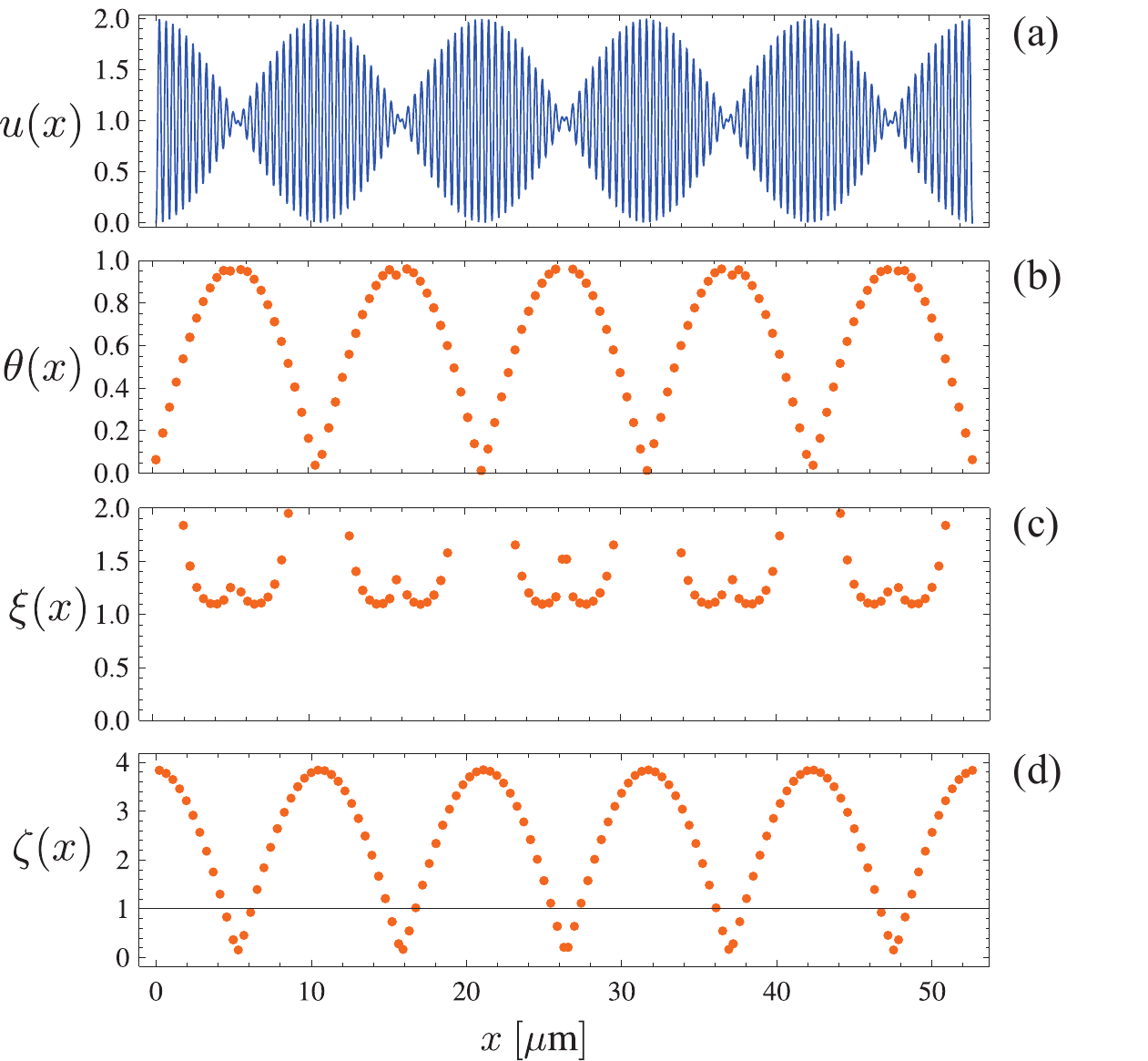}
\caption{(a) Spatial dependence $u(x)$ of the AC Stark potential along the cavity axis for a cavity length $L\simeq 53\mu$m. The two driven modes are at $\lambda_1\simeq 852$nm and $\lambda_2\simeq 888$nm. Their separation is q=5~FSRs. Both the parameter $\theta(\Xabar)$ (b) entering the atom-cavity coupling $g_\at$ and the parameter $\xi(\Xabar)$ (c) entering the atomic dephasing rate $\Gamma_\at$ can be kept close to 1 at potential wells around the points where $\delta k x=n\pi$ with $\delta k=k_1-k_2$ and $n\leq q$. In (d) the parameter $\zeta(x)$ is shown, which can be well around 50\% while $\theta,~\xi\simeq 1$. The resulting loss in trap frequency can be easily compensated for by an increased intracavity amplitude.}
\label{figModes}
\end{centering}
\end{figure}

Let us first discuss briefly how the various requirements on the AC Stark potential generated by the two cavity modes can be met. These requirements are as follows: (i) Above we have assumed for the atom-cavity coupling $g_\at =U_0\alpha\eta\theta(\Xabar)$ and for the diffusion rate $\Gamma_\at=\gamma\frac{g_\at^2}{\Omega_0^2}\xi(\Xabar)$ that both geometrical factors, $\theta(\Xabar)$ and $\xi(\Xabar)=[2 - (4/5)u(\Xabar)]/\theta^2 (\Xabar)$ (for $\Delta m = 0$), can be of order one for a proper choice of the atomic mean position $\Xabar$ along the cavity axis.
Moreover, it is desirable to keep the value of $\zeta(x)$, which enters the atomic trap frequency, as well close to one. (ii)  The two modes have to couple, respectively, to the $D_1$ and $D_2$ lines of the chosen atomic species. For a micro-cavity this implies that the two modes have to be separated by a couple (say $q$) of free spectral ranges (FSRs) only. A typical intensity profile is shown in Fig.~\ref{figModes}a for a mode separation of $q=5$ FSRs.
(iii) The atom has to be located in one of the potential wells determined by $u'(x)=0$. If we take $k_{1(2)}=(k\pm\delta k)/2$ then this condition is equivalent to
\[
\tan(k x)=-\frac{\delta k}{k}\tan(\delta k x).
\]

In Fig.~\ref{figModes}b,~c and d we show exemplarily the values of the parameters $\theta,~\xi$ and $\zeta$ for the possible potential wells, i.e. for the solutions to the last equation. As can be seen the intensity maxima which exhibit the desired properties lie close to points where the cavity modes are almost completely out of phase, that is at points where $\delta k x=n\pi$ for some natural number $n\leq q$.


\subsection{Membrane heating due to laser absorption}
\label{subsec:absorp}

Thermal decoherence depends on the ambient temperature $T$ of the membrane. It can be reduced by precooling the membrane with a cryostat. However, it is important to note that there is a natural lower limit for $T$ which is set by absorption of laser light inside the membrane. The intracavity light hits the membrane in its center, where a fraction $a=P_a/P_c$ of the overall circulating power $P_c=\tfrac{\hbar\omega_\cav c \alpha^2}{L}$ in the two cavity modes is absorbed. If the cavity finesse $\mathcal{F}$ is limited by absorption inside the membrane, we can estimate $a\simeq \tfrac{2\pi}{\mathcal{F}}$. The absorbed power $P_a$ causes an increase of the temperature of the membrane center by $\Delta T \simeq \tfrac{1}{k_B\kappa_\mathrm{th}} P_a$, where $\kappa_\mathrm{th}$ is the thermal link of the membrane center to the membrane supporting frame \cite{Zink04}. $\kappa_\mathrm{th}$ depends on the specific geometry and the material properties and is chosen here such as to have dimensions of Hz. While it is not entirely clear how the resulting inhomogeneous temperature distribution exactly affects the vibrational mode in question, a safe assumption is an increase of the ambient temperature by $\Delta T$.

In \cite{Zink04}, experiments on heat transport inside SiN membranes at cryogenic temperatures were performed. By rescaling the thermal link measured in \cite{Zink04} at a temperature of $\simeq 2$~K to our geometry, $k_B\kappa_\mathrm{th} \simeq 10~\mathrm{nW}/\mathrm{K}$ is obtained. We furthermore use our parameters $P_c=850~\mu$W and $\mathcal{F}=2\times 10^5$, noting that this value for $\mathcal{F}$ is consistent with an imaginary part of the refractive index of the membrane of $\mathrm{Im}(n) \simeq 1\times 10^{-5}$ \cite{wilson_cavity_2009}. With these parameters, we obtain $\Delta T \simeq 2.5$~K. Cryogenic precooling of the membrane frame to $T_0 < \Delta T$ thus allows one to obtain membrane temperatures of the order of $T \simeq \Delta T$.

To gain further insight into the temperature distribution inside the membrane, we simulate heat transport in the membrane by solving the heat equation in 2D with the finite elements method. We assume that the absorbed power $P_a$ is homogeneously distributed over an area $A=\pi w_0^2$ in the membrane center, where $w_0 = 10~\mu$m is the beam waist of the cavity mode, and that the membrane frame is held at a fixed temperature of $T_0=2$~K. At this temperature, the thermal conductivity of SiN is $k_\mathrm{th}=0.05~\mathrm{W}/\mathrm{m}\,\mathrm{K}$ \cite{Zink04}.

\begin{figure}[tbh]
\centering
\includegraphics[width=0.8\columnwidth]{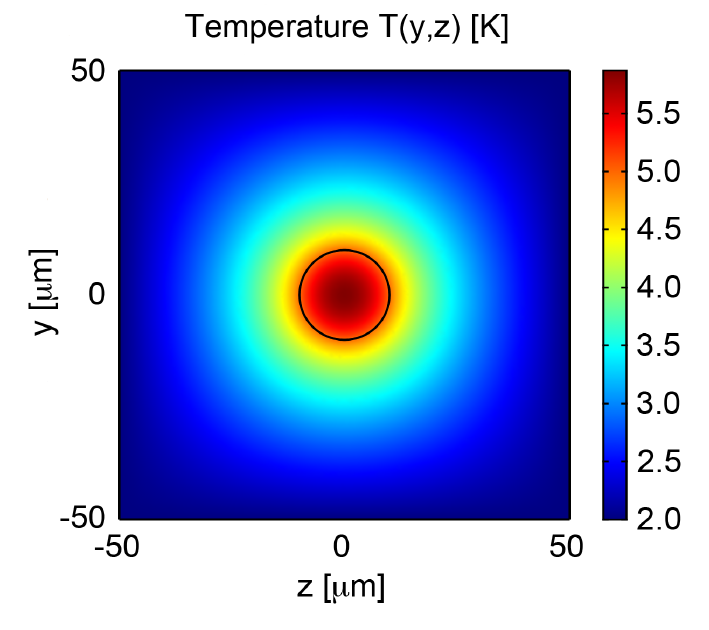}
\caption{Finite elements simulation of membrane heating due to laser absorption. The steady-state temperature distribution $T(y,z)$ in the membrane is shown for the experimental parameters considered in this paper. The laser spot of radius $w_0=10~\mu$m is indicated by the black circle. The membrane frame is held at fixed $T_0 = 2$~K.
\label{fig:membraneheating}
}
\end{figure}

Figure~\ref{fig:membraneheating} shows the steady-state temperature distribution $T(y,z)$ in the membrane obtained from the simulation. The peak temperature in the membrane center is $T(0,0)=5.8$~K. The average temperature obtained by integrating $T(y,z)$ over the membrane cross sectional area is $\bar T = 2.8$~K.
We note that our simulation overestimates the temperature increase, because a constant value for $k_\mathrm{th}$ is used, while in reality $k_\mathrm{th}$ increases rapidly with temperature \cite{Zink04}.
In summary, we conclude that membrane heating due to laser absorption sets a lower limit on the attainable $T$, but for our parameters still allows for cryogenic precooling to $T$ of a few kelvin.

\section{Outlook and Conclusions}
\label{Sec:Disc}

In this paper we have discussed the coupling of the motion of a single atom and a mesoscopic mechanical oscillator, giving rise to a coupled oscillator dynamics. As an outlook, we want to indicate how the present setup could in principle also be used for implementing a Jaynes-Cummings model by coupling the membrane vibrations to the {\it internal} atomic degrees of freedom. Consider an atom with two stable ground states trapped by an external potential inside the cavity. Let both levels be Stark-shifted by the cavity mode, but in \textit{opposite} directions. The coupling of the cavity field quadrature to the two level system is then given by
\[
{\Omega_0^2 \over \delta}\alpha (\hat{a} + \hat{a}^\dag) \sigma_z .
\]
Changing basis and doing the rotating wave approximation, we find an atom-cavity interaction of Jaynes-Cummings form,
\[
\Ga (\hat{a}\sigma_+ + \hat{a}^\dag \sigma_-), \qquad \Ga = \U\alpha .
\]
From this brief derivation we learn that by coupling directly to the internal levels one wins a Lamb-Dicke parameter $\eta \ll 1$ in the coupling constant, compared to coupling to the atomic motion. However, this comes at the prize of an increased dissipation rate, as the spontaneous emission from excited states with rate $\gamma$ translates into ground-state dephasing $(\Gamma_\at/2) {\cal D}[\sigma_z](\rho)$ with rate
\[
\Gamma_\at \sim \gamma p_e
\]
which is a factor $1/\eta^2$ larger, compared to the momentum diffusion. In order to profit from the increased coupling strength, it would thus be crucial to further suppress spontaneous emission by techniques such as e.g. coherent population trapping. Our main point here is however not to declare a winning model class, but rather to point out the various types of interaction which can be implemented with the present setup. Realizing the Jaynes-Cummings model as described above allows for experiments along the lines of microwave CQED \cite{haroche_exploring_2006}.

Overall, our results illustrate that the present setup actually provides a toolbox for engineering various interactions and different types of dynamics, which pave the way towards quantum state engineering of and full quantum control over massive micromechanical systems.

\section*{Acknowledgement}

Support by the Austrian Science Fundation through SFB FOQUS, by the IQOQI, by the European Union through project EuroSQIP, by NIST and NSF, and by the DFG through NIM, SFB631, and the Emmy-Noether program is acknowledged. C.G. is thankful for support from Euroquam Austrian Science Fund project I1 19 N16 CMMC. M.W., K.H., P.Z. and J.Y. thank H.J.K. for hospitality at CalTech.
\appendix

\section{Adiabatic elimination of the cavity modes}\label{App:Elim}

In this section we present details of the
adiabatic elimination of the (independent) cavity modes, starting from the linearized master equation in the interaction picture w.r.t. $H_0$,
\[
\dot{W} = - i[H_{\rm int}(t), W] + \left[ L_\mec + L_\at + \sum_i L_{i} \right] (W),
\]
and further assuming atom and mirror to be on resonance, $\oma = \omm$.
The method we employ here is the projection technique \cite{QuantumNoise} which assumes separation of timescales; that the fast cavity dynamics on the time scale of the interaction allows us to approximate the density operator as $W \simeq \rho_\cav^0 \otimes \rho$ for times $t > 1/\Delta_i$. Here $\rho_\cav^0 = \otimes_i |0\rangle_i \langle 0 |_i$ is the steady-state of the shifted cavity modes, and $\rho={\rm Tr}_\cav (W)$ is the reduced density operator for atom and membrane motion. Furthermore, the influence of the cavity on the membrane-atom dynamics, i.e. the correction in $\rho$ to the free dynamics, is included through second-order perturbation expansion in the interaction term, using $g_i/(\di \pm \omm ) \ll 1$. Finally the expressions are simplified using the Born-Markov approximation. A crucial point is the assumption that the cavity dynamics dominates over the independent dissipation mechanisms of membrane and atom,
$|\Delta \pm \omm| \gg \ga,\Gamma_\mec$, which would otherwise complicate the cavity-mediated dynamics considerably.
The result of the adiabatic elimination is an effective master equation for the atom-membrane system,
\[
\dot{\rho} = \left( L_\mec + L_\at + L_{\rm c-med} \right) (\rho) ,
\]
with cavity-mediated atom-membrane dynamics,
\begin{align*}
& L_{\rm c-med} (\rho) = -
\int_0^\infty d\tau {\rm Tr}_\cav \\
& \left\{
\left[ H_{\rm int}(t) , \left(\otimes_i e^{L_{i}\tau}\right) \left[ H_{\rm int}(t - \tau), \rho_\cav^0 \otimes \rho (t) \right] \right] \right\} \ .
\end{align*}
After tracing over the cavity modes, we find that due to the independence of the cavity modes (only combinations of the form ${\rm Tr}_\cav \{ \aid\ai \rho_\cav^0 \}$ contribute) the cavity-mediated Liouvillian is a sum of contributions from the different modes $i$,
\begin{align*}
& L_{\rm c-med} (\rho) = \sum_i
g_i^2 \int_0^\infty d\tau e^{- (\kappa + i\di ) \tau} \\
& \left[ \left( \Fi(t) + \Fid (t) \right) , \rho (t) \left(\Fi(t-\tau)+\Fid (t-\tau)\right)
\right] + {\rm h.c.}  \ \ .
\end{align*}
Performing the integrals and returning to the lab frame, we find
\begin{eqnarray}
& L_{\rm c-med}(\rho) =
\sum_i \left[ \left( \Fi + \Fid \right)\rho \left( \hmi \Fi + \hpi \Fid \right) + {\rm h.c.} \right] \nonumber \\
& -  \sum_i \left[ \rho \left( \hmi \Fi + \hpi \Fid \right) \left( \Fi + \Fid \right) + {\rm h.c.} \right]
\label{Eq:Lcmed}
\end{eqnarray}
with the constants $h_{\pm,i} = g^2 / [\kappa + i (\di \pm \omm)]$ .

We expect the cavity-mediated dynamics to be described by
\be
 L_{\rm c-med}(\rho) = -i \left[ H_{\at -\mec} , \rho \right] + L_\cav (\rho) ,
\label{Eq:Lwish}
\ee
with cavity-mediated  interaction $H_{\at-\mec}$ (including corrections to the free dynamics) and cavity-mediated decay $L_\cav (\rho)$. In order to compare the expectation (\ref{Eq:Lwish}) with the result (\ref{Eq:Lcmed}), we split the second line of (\ref{Eq:Lcmed}) into commutator and anticommutator parts,
\[
- \left( \rho A + {\rm h.c.} \right) =  -i \left[ {i \over 2}\left( A - A^\dag\right) ,\rho\right]
- \left\{ {1\over 2}\left(A + A^\dag \right) ,\rho\right\}_+ ,
\]
with $A = \sum_i \left( \hmi \Fi + \hpi \Fid \right) \left( \Fi + \Fid \right)$.
Reading off from the commutator that $H_{\at -\mec} = (i / 2)( A - A^\dag )$, we find the cavity-mediated coherent dynamics,
\begin{align}
H_{\at-\mec} =
{i \over 2}\sum_i \left[\left( \hmi \Fi + \hpi \Fid \right) \left( \Fi + \Fid \right)  - {\rm h.c.} \right].
\label{Eq:HCmed}
\end{align}
The anti-commutator from the second line of (\ref{Eq:Lcmed}) combined with the sandwich terms in the first line of (\ref{Eq:Lcmed}) describes correlated decay of membrane and atomic motion through the cavity,
\begin{eqnarray}
& L_\cav(\rho) = \sum_i
{1 \over 2}\Big[ 2 \left( \Fi + \Fid \right) \rho \left( \hmi\Fi + \hpi \Fid \right) \nonumber \\
& - \left\{ \left( \hmi\Fi + \hpi \Fid \right)\left( \Fi + \Fid \right) , \rho \right\}_+ \Big]
+ {\rm h.c.}.
\nonumber \\
\label{Eq:DissCmed}
\end{eqnarray}
When written in this form the decay is not manifestly on Lindblad form ${\cal D}[a](\rho) = 2 a \rho a^\dag - \{ a^\dag a , \rho\}_+$. However, it is possible to diagonalize the Liouvillian (\ref{Eq:DissCmed}) and write it as a sum of 2 independent jump processes ($k=1,2$) for each cavity mode $i$,
\[
L_\cav(\rho) = \sum_{i,k} {\gamma^{(i)}_{k}\over 2} {\cal D}\left[J^{(i)}_{k} \right](\rho),
\]
with the jump operators $J^{(i)}_{k}$,
\[
J^{(i)}_{k} =  \left(\vec{m}^{(i)}_k \right)^T
\left(
\begin{array}{c}
\Fi \\
 \Fid
\end{array}
 \right),
\]
which are described by the eigenvectors $\vec{m}^{(i)}_k$ of the $2\times 2$ matrices $\hat{M}_i $,
\be
\hat{M}_i = \left(
\begin{array}{cc}
2 {\rm Re} \{ \hpi \} & \left( \hmi + \hpi^* \right) \\
\left( \hmi + \hpi^* \right)^* & 2 {\rm Re} \{ \hmi \}
\end{array}
\right).
\label{Eq:Mmatrix}
\ee
The corresponding decay rates $\gamma^{(i)}_{k}$ are given by the eigenvalues of $\hat{M}_i$.
However in view of the fast rotations of terms of the form $\Fi\rho\Fi$ for $\omm \gg \gamma^{(i)}_{k}$, we perform the rotating wave approximation (RWA), which here corresponds to omitting the off-diagonal terms in $\hat{M}_i$, leading to (\ref{Eq:LcRWA}).

\section{Time evolution and state transfer}
\label{sec:timeevol}

The reduced atom-membrane master equation (\ref{Eq:MEeff}) can be written on the general form
\be
\dot\rho = -i \left[ \vec{\rm R}^T \hat{\rm H}\vec{\rm R}, \rho \right]
+ \sum_k {\gamma_k \over 2} {\cal D} \left[\vec{\rm L}_k^T \vec{\rm R}\right](\rho)
\label{Eq:MEgen}
\ee
with the matrix $\hat{\rm H}$ describing the Hamiltonian dynamics and the vectors $\vec{\rm L}_k$ describing the jump operators of the dissipation processes. At this point it is convenient to switch to the dimensionless $\{X,P\}$ language, and thereby introduce the basis vector $\vec{\rm R} = [X_\mec,P_\mec,X_\at,P_\at]^T$ with
$X_\at = (\aat + \aad)/\sqrt{2}$ and $X_\mec = (\am + \amd)/\sqrt{2}$ and the commutation relations collected in a matrix $\hat\sigma$,
\[
\left[{\rm R}_i , {\rm R}_j \right] = i \sigma_{ij},
\qquad
\hat{\sigma} =
\left[
\begin{array}{cc | cc}
0 & 1 & & \\
-1 & 0 & & \\
\hline
 & & 0 & 1 \\
 & & -1 & 0
\end{array}
\right] .
\]
The time evolution of a Gaussian state is fully described by its covariance matrix $\hat\gamma (t)$ and displacement vector $\vec{\rm d}(t)$, which are defined as the first and second moments, respectively,
\[
\vec{\rm d} = \langle \vec{\rm R} \rangle,
\qquad
\hat\gamma_{ij} = {1 \over 2}\langle {\rm R}_i {\rm R}_j + {\rm R}_j {\rm R}_i \rangle -  \langle {\rm R}_i \rangle \langle {\rm R}_j \rangle .
\]
From the general master equation (\ref{Eq:MEgen}) one can derive the equations of motion for the moments,
\[
\dot{\vec{\rm d}}(t) = \hat{\rm Q} \vec{\rm d}(t)
, \qquad
\dot{\hat\gamma}(t) = \hat{\rm Q}\hat\gamma(t) + \hat\gamma(t) \hat{\rm Q}^T + \hat{\rm N} ,
\]
with the matrices $\hat{\rm Q}$ and $\hat{\rm N}$ given by,
\be
\hat{\rm Q} = 2 \hat{\sigma} \left( \hat{\rm H} + {\rm Im } \{ \hat\Gamma \} \right),
\qquad \hat{\rm N} = 2 \hat{\sigma}\left[{\rm Re } \{ \hat\Gamma \}\right]\hat{\sigma}^T .
\label{Eq:QN}
\ee
Here the matrix $\hat\Gamma$ collects the information about the various dissipation channels,
\[
\hat\Gamma_{mn} = \sum_k {\gamma_k \over 2} \left( {\rm L}_{k,m}^* {\rm L}_{k,n}\right).
\]
Using this technique, we can solve for the time evolution of Gaussian states. In particular the solution for the time evolution of the covariance matrix reads,
\be
\hat\gamma (t) = e^{\hat{\rm Q}t}\hat\gamma (0) e^{\hat{\rm Q}^T t} +
\int_0^t d\tau e^{\hat{\rm Q}(t-\tau)} \hat{\rm N}e^{\hat{\rm Q}^T (t-\tau)} .
\label{Eq:eqmCovM}
\ee


We now go on to present the analytical results for state transfer which will be used in section \ref{sec:cohevol}, based on the calculations in App. \ref{sec:timeevol}. Along the lines of section \ref{sec:cohevol} we consider the symmetric two-mode setup with  $\Gm = \Ga$ $(= g/\sqrt{2})$, assuming the rotating wave approximation $\omm \gg \Geff, \Gamma_\at, \gamma_k^{(i)}$ and the large detuning regime, $|\Delta| \gg \omm, \kappa$, and neglecting the difference in membrane heating and cooling, $1/\nm \ll 1$. For this special case, the jump operators form cooling/heating pairs with equal rate $\Gamma$; if a cooling channel is described by some vector $\vec{\rm L}$, then the corresponding heating channel is described by its complex conjugate $\left(\vec{\rm L}\right)^*$,
\[
\dot\rho \sim {\Gamma \over 2}\left(  {\cal D} \left[\vec{\rm L}^T \vec{\rm R} \right](\rho)+
{\cal D} \left[\left(\vec{\rm L}^T\right)^* \vec{\rm R} \right](\rho) \right).
\]
Thus, in this particular case, the sum of the respective contributions to the dissipation matrix $\hat\Gamma$ from the cooling and the heating processes is by definition real,
\[
{\rm Im } \{ \hat\Gamma \} = 0 .
\]
Hence dissipation does not enter the matrix $\hat{\rm Q}$ (\ref{Eq:QN}). Consequently the time evolution of the displacement vector $\vec{\rm d}(t)$ is completely coherent, and the effect of dissipation on the state can only be seen in the evolution of the covariance matrix $\hat\gamma (t)$.

\subsection{Transfer of coherent or squeezed states}

The thermalization of a coherent state during the atom-membrane interaction is here studied by solving the equations of motion (\ref{Eq:eqmCovM}) for the covariance matrix $\hat\gamma (t)$. For an initial coherent state,
\[
\hat\gamma (0)
= {1 \over 2}\left(
\begin{array}{c|c}
{\bf 1}  & \\
\hline
 & {\bf 1}
\end{array}
\right)
\]
we find the following evolution
\[
\hat\gamma \left(t \right)=
{1 \over 2}\left[ 1 + 2\bar{n}(t)\right]\left(
\begin{array}{c|c}
{\bf 1}  & \\
\hline
 & {\bf 1}
\end{array}
\right)
+
\hat\gamma_{\rm corr}(t)
\]
describing a thermal state with dissipation-induced population
\[
\bar{n}(t) = {1 \over 2}\left( 2 \Gamma_\cav + \Gamma_\mec + \ga\right)t .
\]
The second part of the covariance matrix, $\hat\gamma_{\rm corr}$, describes oscillating atom-membrane correlations which are only present when atom and membrane dissipate with different rates,
\[
\hat\gamma_{\rm corr} ={\Gamma_\mec - \ga \over 2\Geff}
\left(
\begin{array}{cc | cc}
s_1(t)&  &  & s_2(t) \\
 &  s_1(t) & -s_2(t) & \\
 \hline
 & -s_2(t)  & -s_1(t) & \\
s_2(t) &  &  &  - s_1(t)
\end{array}
\right),
\]
with $s_1(t) = \sin[2\Geff t]$ and $s_2(t) = \sin^2[\Geff t]$. Due to thermalization the fidelity of the state transfer decreases with time from $F(0) = 1$ according to
\be
F(t) = {1 \over \sqrt{{\rm det}\left(\hat\gamma_\mec (t)+ \hat\gamma_\at (0)\right)}}
\label{Eq:fidgeneral}
\ee
as presented for a state swap ($t=\pi/2\Geff$) in Eq. (\ref{Eq:fid}) and Fig. \ref{figCoh}(c).
Note that this measure does not take into account the (coherent) rotation of the displacement vector.

For the transfer of a squeezed state we see a similar thermalization, as is clear from Fig. \ref{figSqueez}, where we study swap of a state with an initial atomic quadrature squeezing of $e^{-2}$
\[
\hat\gamma_\at \left(0 \right)=
{1 \over 2}\left(
\begin{array}{cc}
e^{-2} & \\
& e^2
\end{array}
\right).
\]
In this case the covariance matrix evolves according to
\be
\hat\gamma_\mec^\prime \left(t= {\pi \over 2\Geff} \right)=
{1 \over 2}
\left(
\begin{array}{cc}
e^{-2} + 2\bar{n}_{\rm swap} & \\
& e^{2} + 2\bar{n}_{\rm swap}
\end{array}
\right).
\label{Eq:covsq}
\ee
Here $\hat\gamma_\mec^\prime(t)$ is the diagonalized membrane covariance matrix, and the thermal population is given by
\be
\bar{n}_{\rm swap} \equiv \bar{n} \left({\pi \over 2\Geff} \right) =
{\pi \over 4G}\left( 2 \Gamma_\cav + \Gamma_\mec + \ga\right).
\label{Eq:nswap}
\ee

\subsection{Transfer of a Fock state}
\label{subsec:Fock}

Our figure of merit for the thermalization of a Fock state during state transfer to the membrane, is the negativity of the membrane Wigner function $w_\mec (\beta,\beta^*,t)$ at the origin $\beta = \beta^* = 0$ relative to the (absolute) negativity of the Fock state Wigner function $w_{\rm F}(0,0)$,
\be
N_{\rm w} (t) \equiv {w_\mec(0,0,t)  \over |w_{\rm F} (0,0)| }
= { 1 \over 4 \pi} \int d^2 \xi_\mec \chi_\mec  (\xi_\mec,t) .
\label{Eq:neg}
\ee
Here $\chi_\mec  (\xi_\mec,t)$ is the characteristic function of the membrane state, which is derived from the characteristic function of the total atom-membrane state,
\[
\chi_\mec  (\xi_\mec,t) =
\chi_{\mec,\at}  (\xi_\mec,\xi_\at = 0, t).
\]
In order to evaluate this expression we need the time evolution of the full characteristic function, which is defined in terms of the density operator $\rho_{\mec,\at} (t)$,
\[
\chi_{\mec,\at}  (\xi_\mec,\xi_\at, t) = {\rm Tr} \left\{ \rho_{\mec,\at} (t) D_\mec(\xi_\mec)D_\at(\xi_\at) \right\}
\]
with $D(\xi) = \exp\left[ \xi \hat{a}^\dag - \xi^* \hat{a} \right]$ the displacement operator. Let us first consider the coherent membrane-atom state, for which the time evolution is described by the covariance matrix $\hat\gamma(t)$ and the displacement vector $\vec{\rm d}(t)$,
\[
\chi_{\alpha,\beta}  (\xi_\mec,\xi_\at,t) =
\exp \left[ -{1 \over 2}
\vec{\xi}^T
 \hat\gamma (t)
\vec{\xi}
+
i \vec{\rm d}(t) \vec{\xi}
\right]
\]
with
\[
\vec\xi =
\sqrt{2}\hat\sigma
\left(
\begin{array}{c}
{\rm Re} (\xi_{\mec}) \\
{\rm Im} (\xi_{\mec}) \\
{\rm Re} (\xi_{\at}) \\
{\rm Im} (\xi_{\at})
\end{array}
\right).
\]
Our next attempt is to use the result for coherent states to simplify the description of the evolution of a Fock state; here we focus on the state $|1\rangle$. We note that by expressing the Fock state in terms of coherent states,
\[
|1\rangle = \partial_\alpha \left( e^{|\alpha|^2/2} |\alpha \rangle \right)_{\alpha = 0},
\]
the characteristic function $\chi_1 (\xi)$ for $|1\rangle$ can be written in terms of corresponding functions for coherent states,
\begin{equation}
\chi_1 (\xi)
= \partial^2_{\alpha,\alpha^*} \left( e^{\alpha\alpha^*} \chi_\alpha (\xi) \right)_{\alpha = 0}.
\label{chi1}
\end{equation}
Consequently, for an initial density matrix $\rho(0) = |0\rangle \langle 0|_\mec|1 \rangle\langle 1|_\at$ we derive the initial characteristic function $\chi  (\xi_\mec,\xi_\at,0)$,
\begin{align}
& \chi  (\xi_\mec,\xi_\at,0) = \nonumber \\
& \partial^2_{\alpha,\alpha^*} \left( e^{\alpha\alpha^*}
\exp \left[ -{1 \over 2}
\vec\xi^T
\hat\gamma (0)
\vec\xi +
i \vec{\rm d}(0)
\vec\xi \right]
\right)_{\alpha = 0}
\label{chi-t0}
\end{align}
with initial conditions
\[
\hat{\gamma}(0) = {1 \over 2} \left(
\begin{array}{c|c}
{\bf 1} & 0 \\
\hline
0 & {\bf 1}
\end{array}
\right),
\qquad
\vec{\rm d}(0) =
\sqrt{2}\left(
\begin{array}{c}
0 \\
0 \\
{\rm Re} (\alpha)\\
{\rm Im} (\alpha)
\end{array}
\right) .
\]
Using the linear properties of the time evolution map $\epsilon_t $,
\[
\rho(t) = \epsilon_t  (\rho(0))
\]
we find that the time evolution of the full characteristic function is given by
\begin{align*}
& \chi  (\xi_\mec,\xi_\at,t) = \nonumber \\
& \partial^2_{\alpha,\alpha^*} \left( e^{\alpha\alpha^*}
\exp \left[ -{1 \over 2}
\vec\xi^T
\hat\gamma (t)
\vec\xi +
i \vec{\rm d}(t)
\vec\xi \right]
\right)_{\alpha = 0} ,
\end{align*}
from which we derive the membrane Wigner function negativity (\ref{Eq:neg}); for this specific case it reads,
\be
N_{\rm w} (t) =
{2\bar{n}(t) + \cos[2Gt] + \sin[2Gt](\Gamma_\mec - \ga)/2G
\over \left( 1 + 2\bar{n}(t) + \sin[2Gt](\Gamma_\mec - \ga)/2G\right)^2}.
\label{Eq:FockNeg}
\ee


\end{document}